\documentclass[a4paper,11pt]{article}

\usepackage{jinstpub} 
\usepackage{placeins}
\usepackage{graphicx}
\usepackage{siunitx}

\usepackage{lineno}

\title{Design, characterization and installation of the NEXT-100 cathode and electroluminescence regions}
\author[1,a]{K.~Mistry\note[a]{Corresponding Author},}
\author[2,a]{L.~Rogers,}
\author[1]{B.J.P.~Jones,}
\author[1]{B.~Munson,}
\author[1]{L.~Norman,}
\author[3,1]{D.~Oliver,}
\author[1]{S.~Pingulkar,}
\author[1]{M.~Rodriguez-Tiscareno,}
\author[1]{K.~Silva,}
\author[1]{K.~Stogsdill,}
% ...
\affiliation[1]{
Department of Physics, University of Texas at Arlington, Arlington, TX 76019, USA}
\affiliation[2]{
Argonne National Laboratory, Argonne, IL 60439, USA}
\affiliation[3]{
University of Texas at El Paso, 500 W. University Ave., El Paso, Texas 79968, USA.}
% ...

\emailAdd{krishan.mistry@uta.edu}
\emailAdd{rogersl@anl.gov}

\abstract{NEXT-100 is currently being constructed at the Laboratorio Subterr\'aneo de Canfranc in the Spanish Pyrenees and will search for neutrinoless double beta decay using a high-pressure gaseous time projection chamber (TPC) with 100 kg of xenon. Charge amplification is carried out via electroluminescence (EL) which is the process of accelerating electrons in a high electric field region causing secondary scintillation of the medium proportional to the initial charge. The NEXT-100 EL and cathode regions are made from tensioned hexagonal meshes of 1~m diameter. This paper describes the design, characterization, and installation of these parts for NEXT-100. Simulations of the electric field are performed to model the drift and amplification of ionization electrons produced in the detector under various EL region alignments and rotations. Measurements of the electrostatic breakdown voltage in air characterize performance under high voltage conditions and identify breakdown points. The electrostatic deflection of the mesh is quantified and fit to a first-principles mechanical model.  Measurements were performed with both a standalone test EL region and with the NEXT-100 EL region before its installation in the detector. Finally, we describe the parts as installed in NEXT-100, following their deployment in Summer 2023.}

\begin{document}

\maketitle

\flushbottom
\section{Introduction }
\label{sec:intro}

Whether or not the neutrino is a Majorana particle is one of the leading questions in fundamental experimental  physics, which may elucidate the origin of the matter-antimatter asymmetry in the universe. The most practical known way to answer this question is by searching for a process called neutrinoless double beta decay ($0\nu\beta\beta$)~\cite{snowmasswhitepaper}. High-pressure gas time projection chambers are a compelling technology for such an experiment due to their mm-scale topological reconstruction, sub-percent energy resolution, and versatility. The Neutrino Experiment with a Xenon TPC (NEXT) is an experimental program based on this technology with xenon enriched in $^{136}$Xe serving as both the source and detection medium. The latest experiment in this program to finish operation, NEXT-White \cite{NEXT:2018rgj}, has demonstrated (with 3.5 kg of xenon) superb energy resolution at 1\% FWHM at 2.5~MeV \cite{NEXT:2019qbo}, measured the $2\nu\beta\beta$ half-life \cite{NEXT:2021dqj} and searched for $0\nu\beta\beta$ \cite{NEXT:2023jsn}. The forthcoming NEXT-100~\cite{next100CDR} experiment is currently under construction and will search for $0\nu\beta\beta$ with 100~kg of enriched xenon (91\% $^{136}$Xe) with sensitivity to a half-life of 6$\times$10$^{25}$ years at the 90\% confidence level after three years of operation \cite{Martin-Albo:2015rhw}. 

\begin{figure}[b!]
\centering
\includegraphics[width=\textwidth]{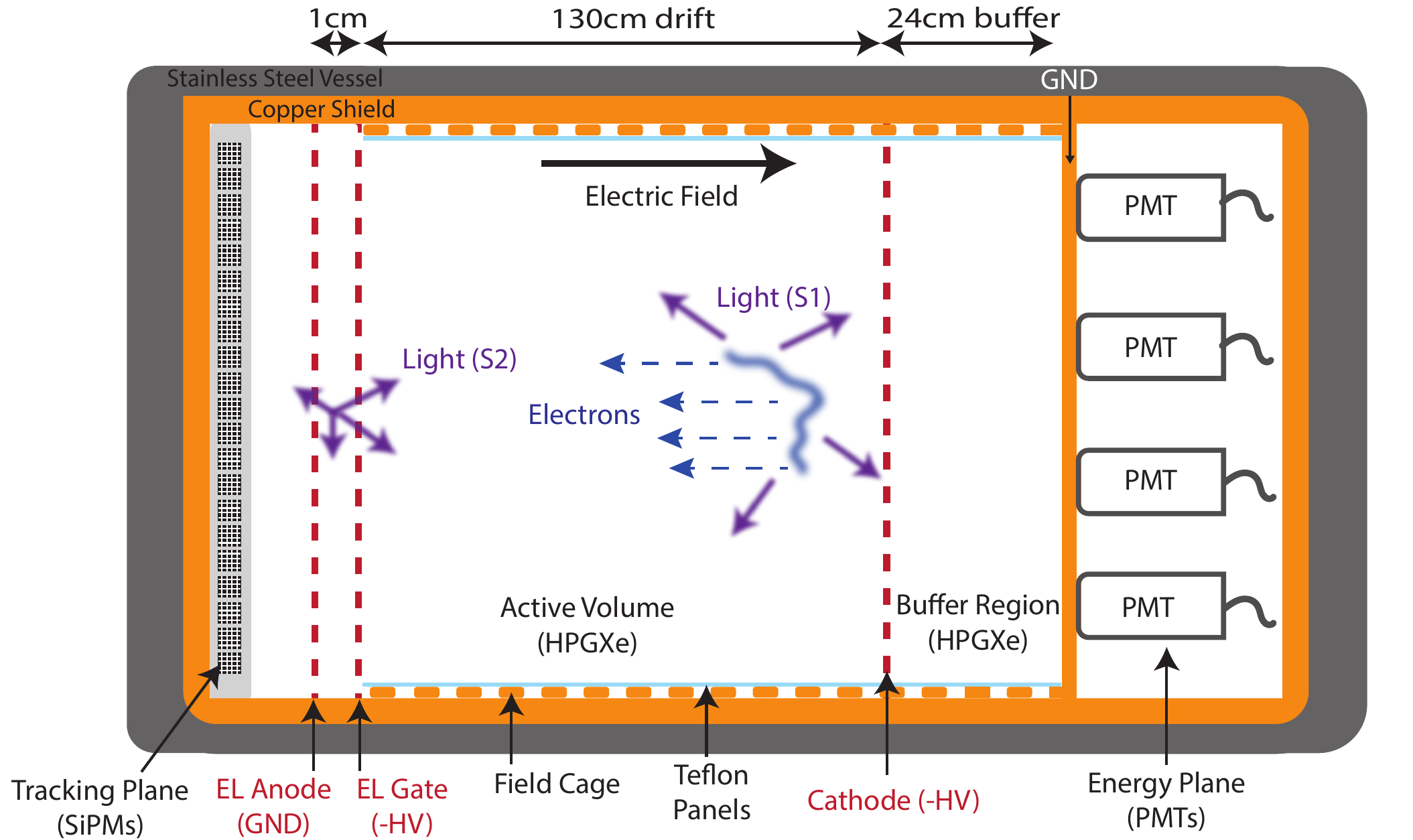}
\caption{\label{fig:next_tpc} A schematic of the main components of the NEXT-100 TPC. Charged particles produced inside the active volume (cylinder of 1~m diameter and 1.3~m length) ionize and excite the xenon producing the S1 signal. Ionization electrons are subsequently drifted to and amplified in the EL region to produce the S2 light. An array of 60 Hamamatsu R11410-10 PMTs and 3584 Hamamatsu SiPMs located at the end-caps are used to measure the energy and provide tracking information from the light produced. The PMTs are separated from the high-pressure gas by sapphire windows. The buffer region steps down the high voltage from the cathode to the energy plane while keeping the electric field below the threshold for electroluminescence. The EL and cathode presented in this paper are highlighted in red.}
\end{figure}

\begin{figure}[t]
\centering
\includegraphics[width=0.282\textwidth]{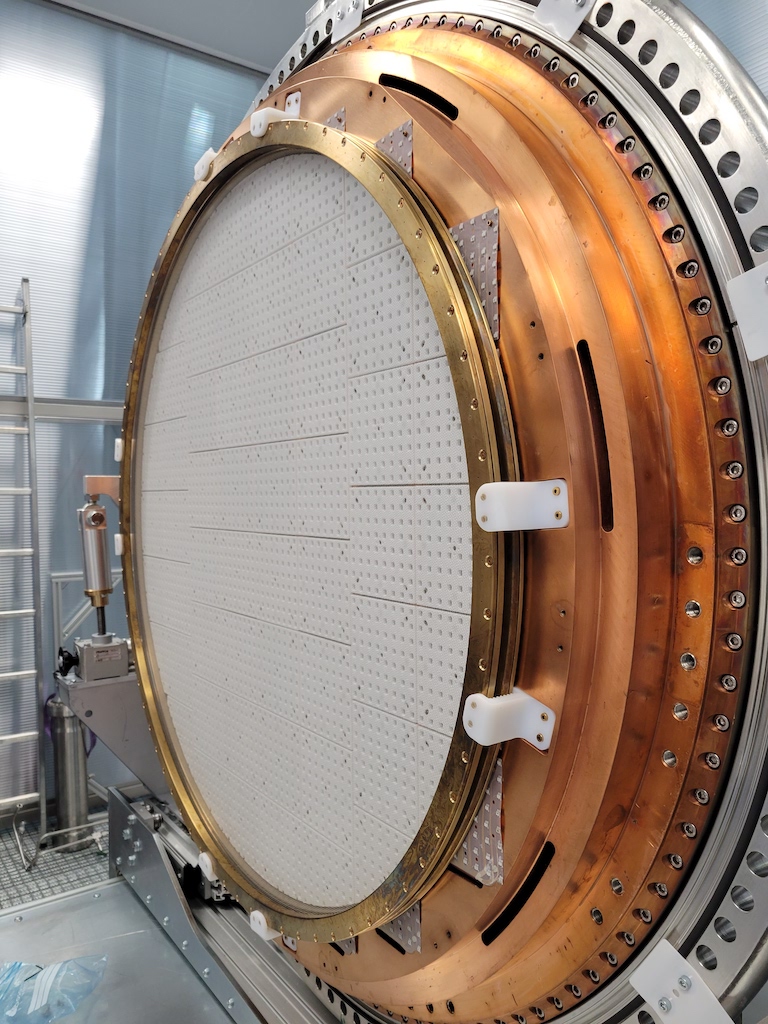}
\includegraphics[width=0.5\textwidth]{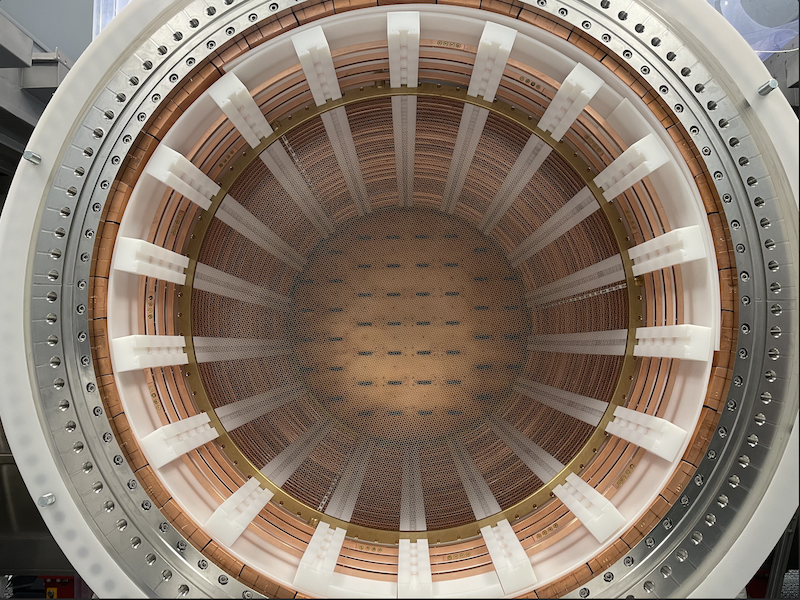}
\caption{\label{fig:el_and_cathode} Left: image of the NEXT-100 EL region attached to the copper tracking plane. The white region is a Teflon mask which improves light collection to the SiPMs which sit just behind. Right: image of the NEXT-100 cathode inside the field cage. Both the EL and cathode are golden rings with a transparent mesh.  }
\end{figure}

The TPC design of NEXT-100 is shown in Fig.~\ref{fig:next_tpc}. As charged particles pass through the drift (active) region of the detector, they ionize and excite the xenon gas. Prompt scintillation light is produced from the excitation of the xenon gas by the primary interaction, known as the S1 signal, and is used as a trigger to determine the interaction time of the event. For the ionization electrons, an electric field of 500~V/cm, generated by applying a large negative potential to the cathode, sweeps them towards the electroluminescence (EL) region which has a higher electric field of 18~kV/cm. After the ionization charges enter this region, they are accelerated and gain sufficient energy to excite (but not ionize) the xenon gas. This leads to the production of approximately 10$^2$ -- 10$^3$ scintillation photons per electron, known as the S2 signal. The photons produced from the EL are collected by silicon photomultipliers (SiPM) and photomultiplier tubes (PMTs) after being wavelength-shifted to the visible spectrum by tetraphenyl butadiene coated onto various surfaces inside of the TPC including the inner Teflon panels, sapphire windows in front of the PMTs, and SiPMs. This information is used to provide topological information and precision calorimetry. Pictures of the EL and cathode regions in NEXT-100 are shown in Fig.~\ref{fig:el_and_cathode}.

The EL region produces S2 light with proportional gain and minimal fluctuations. It has to meet exacting mechanical and electrical design specifications in order to achieve the uniformity and gain required for high-quality topology reconstruction and energy measurement. The NEXT experiment requires percent-level or lower energy resolution for mitigating the contamination of the $2\nu\beta\beta$ background in the signal region as well as reducing the impact of gamma rays that produce energy deposits near the energy range of interest. The electric field profile between the meshes governs the path an electron will take as it crosses the gap which directly affects the light yield and uniformity. In addition, with such high electric fields, the electrostatic force is large enough to cause the two surfaces of the EL region to deflect towards one another, potentially compromising field uniformity and high voltage stability. Defects and sharp edges on the mesh must be avoided, in order to ensure suitable operation at voltages of several tens of kilovolts without electrical sparking or  field emission of electrons. The EL region must have high optical transparency to ensure that a suitable fraction of the light reaches the photodetectors. To ensure these specifications are met, significant mechanical challenges need to be overcome while using  as little material as possible due to radiopurity constraints.  

The design selected to meet these requirements for NEXT-100 consists of two parallel photochemical etched hexagonal meshes tensioned on a metal frame.  In this paper, we describe the design, characterization, and installation of the EL and cathode regions in the NEXT-100 detector. The design and construction of the EL region are first described in Sec.~\ref{sec:design}. Measurements of the mesh robustness and electrical breakdown voltage in air of the EL region are studied in Sec.~\ref{sec:breakdown}. In Sec.~\ref{sec:deflection} we perform measurements of the electrostatic deflection, achieving 8~\si{\micro\meter} precision. Measurements were also performed on the NEXT-100 EL region before it was installed. We report comparisons of these measurements to an analytical model developed that can be used to calculate the deflection at various electric fields and estimate the tension of the meshes with and without a support post. Detailed simulations of the light and field response of the EL and cathode regions using a {\tt COMSOL} \cite{dickinson2014comsol} and {\tt Garfield++} \cite{garfieldplusplus} simulation are presented in Sec.~\ref{sec:simulations}. Finally, a brief report of the installation of the electroluminescent and cathode regions is given in Sec.~\ref{Sec:installation}.

\section{Design and construction of the EL and cathode region}\label{sec:design}

The use of stretched meshes has a wide precedent in xenon TPC experiments and test stands~\cite{Byrnes_2023,baudis2020first,Stephenson_2015,Baur_2023,PhysRevD.99.103024,Lin_2014,APRILE2012573,jorg2022characterization,Jewell_2018,haselschwardt2023measurement,Baudis_2021}.  The challenges associated with the NEXT application are the very large diameter of the detector ($\sim$m) coupled with the high voltage and small gap size required for a large gaseous EL TPC.

Of the three required meshes to delineate the uniform-field regions in NEXT-100, the cathode requirements are the least strict, because it resides in a region of relatively low local electric field.  This means that the expected mesh deflection under electrostatic forces is not a limiting design factor. The cathode design for NEXT-100 consists of a 13.5 mm thick silicon bronze frame with a hexagonal mesh with 5~mm inner diameter hexagons tensioned in between two rings.

\begin{figure}[t]
\centering
\includegraphics[width=\textwidth]{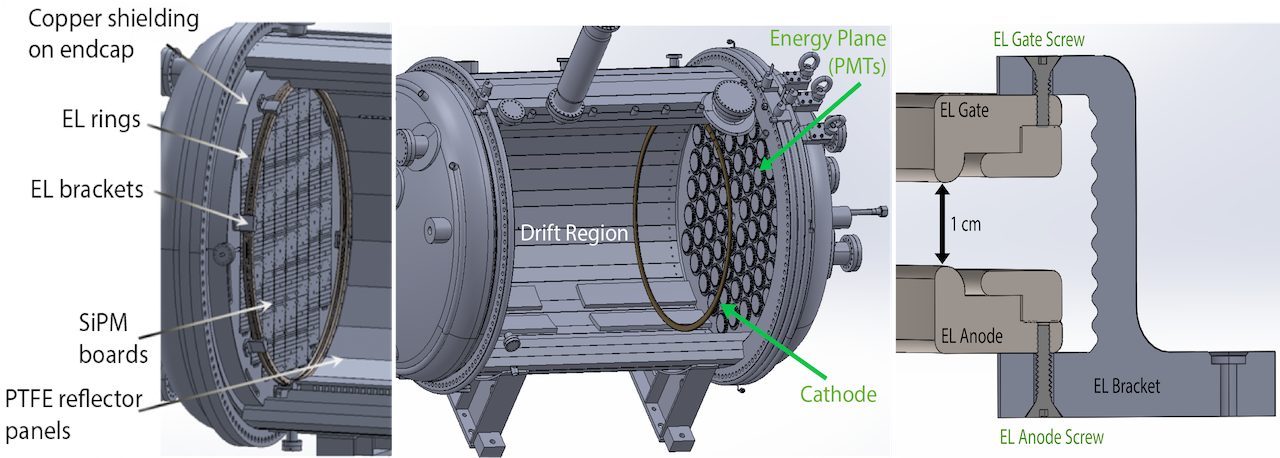}
\caption{\label{fig:NEXT100_EL_Cathode} Left: a side view schematic of the EL region mounted onto the copper end cap at the tracking plane. The EL region sits right in front of the SiPM tracking plane. Middle: view of the cathode inside the detector. The field cage and reflector panels have been omitted from the schematic for visibility. Right: a schematic for how the EL bracket is connected to the EL frames. }
\end{figure}

The EL region is more challenging due to the large fields required for EL gain and the small EL gap width of 1~cm.  It consists of two silicon bronze frames that hold two tensioned meshes with 2.5~mm inner diameter hexagons facing parallel to each other. For the EL region, a large voltage is applied to the ``gate'', i.e. the mesh that the ionization electrons pass through to enter the EL region. The anode is connected to ground. The frames are joined together using eight high density polyethylene (HDPE) brackets located around the circumference which space them with a gap distance of 9.7$\pm$0.15 mm.  These parts are depicted in Fig.~\ref{fig:NEXT100_EL_Cathode}.

\subsection{Photoetched mesh}
\begin{figure}[t]
\centering
\includegraphics[width=0.4\textwidth]{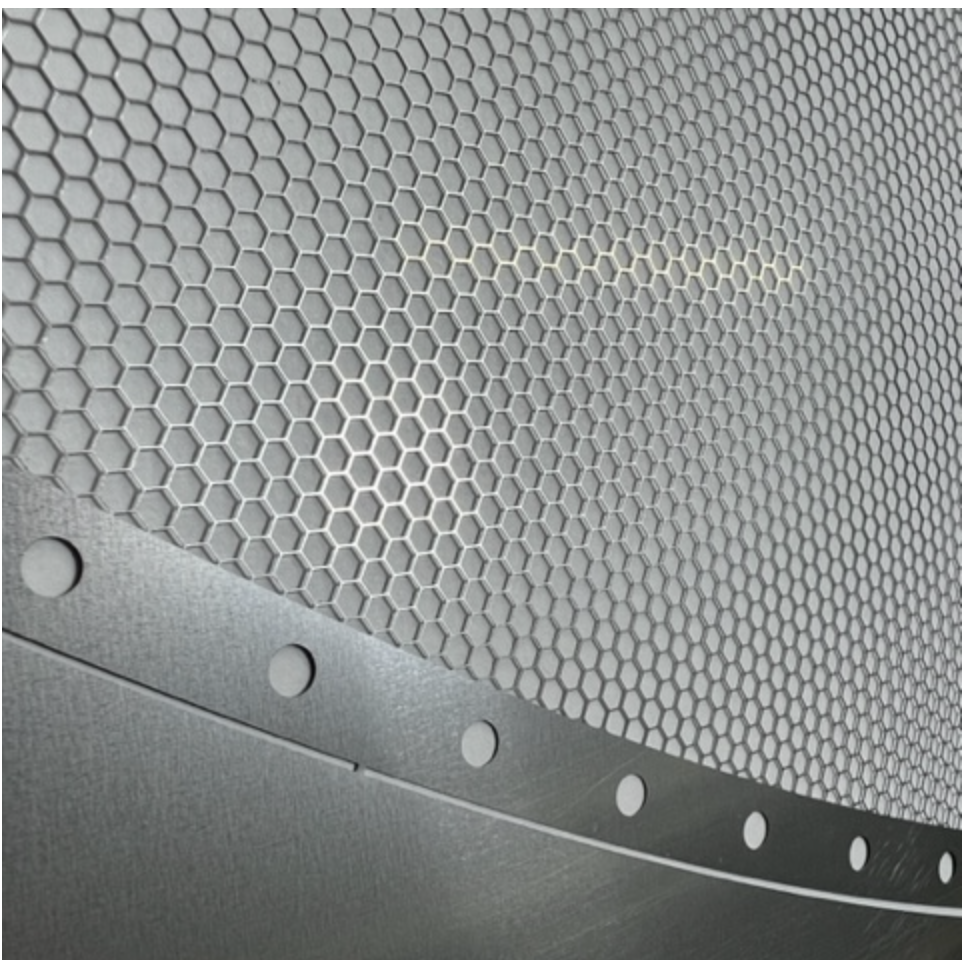}
\caption{\label{fig:EL_Mesh} Image of the NEXT-100 mesh with the 9~mm border and dowel pin holes. The mesh is etched from a single sheet of stainless steel and is held via small tabs. After removal of the mesh from the tabs, the sharp points of these tabs do not pose a breakdown risk as they are located inside the main ring frames. }
\end{figure}

The EL meshes, as shown in Fig.~\ref{fig:EL_Mesh}, are made from a 127~\si{\micro\meter} thick stainless steel (SS) sheet etched with a hexagonal pattern of 2.5 mm inner diameter hexagons and 127~\si{\micro\meter} wire widths. The cathode has a similar design and wire size but with 5~mm inner diameter hexagons. The meshes were manufactured with photochemical etching by PCM Products, Inc. \cite{pcmproducts}. This process is able to achieve tight tolerances down to 5~\si{\micro\meter} depending on the thickness of the material, which can range from 0.5~\si{\micro\meter} to 3 mm. The company has some of the largest etching machines in the USA and is able to etch components as wide as 1.5~m. The advantage of etched hexagonal rather than woven square or wire mesh geometry primarily resides in the frame loading properties. With hexagonal loading, the outer metal frame is not expected to buckle through a ``saddle-shape'' distortion, a common problem for thin frames loaded on one or two perpendicular axes. A further advantage of an etched mesh is that any break does not lead to a loose wire that may short to other surfaces. A fine surface finish is also achievable, giving favorable high-voltage properties.  

While the final design choice for the NEXT-100 hexagonal meshes includes photochemically etched meshes from a single sheet of stainless steel, alternative methods of manufacture were also explored. Previous attempts to manufacture meshes at this scale were attempted with alternative vendors who were not able to etch at the $>$1~m scale by combining partial semi-circular meshes into a full circle via various joining methods. We found that the meshes were too thin and sparse for most welding techniques to work without melting the material. Spot welding resulted in localized points where the tension would be unevenly distributed and resulted in the mesh failing when stretched. Soldering was more robust, with silver-based solder performing the best by achieving the desired tension but then began to fail slowly over time. Thus, we considered single full-sized photoetched meshes as the only viable use case of the photoetching technique for our application. 

The choice of hexagon size was driven by simulations of the electric field uniformity and is described in Sec.~\ref{sec:comsol}. Small hexagons are advantageous in terms of field uniformity and mesh robustness so 2.5~mm meshes were chosen for the EL region (90\% transparency). Larger holes are preferable to maximize optical transparency and optimize light collection efficiency so in the case of the cathode, 5~mm hexagons were chosen (95\% transparency). 

Several iterations of meshes at the scale of NEXT-100 (diameter 1 m) were produced in order to develop parts suitable for installation in the experiment. Initial meshes manufactured included a post-etching process. This includes an additional acid etching step to reduce sharp edges after the primary etching process. However for these meshes, the etching uniformity at scale was compromised with excessive numbers of broken wires and lands (identified with a visual inspection), and poor robustness under tensioning. Further iterations without post-etching significantly improved the quality of the mesh with only a few broken lands on each mesh. Studies on the impact of mesh defects on the electrical breakdown strength were made, and as described in Sec.~\ref{sec:breakdown}, the non-post-etched meshes perform to specification. The best quality of meshes, without breaks but with a few small surface imperfections, was achieved when the manufacturer used a thicker photoresist; this was used for the final meshes used in NEXT-100. 

\subsection{Mesh support and tensioning frames}\label{sec:elframe}

Each EL frame consists of two parts: a base ring for fixing the mesh and a tensioning ring to stretch it. The choice of material for the rings was silicon bronze over stainless steel due to its mechanical strength and radiopurity (described in Sec.~\ref{sec:radioassay}). 

\begin{figure}[t]
\centering
\includegraphics[width=\textwidth]{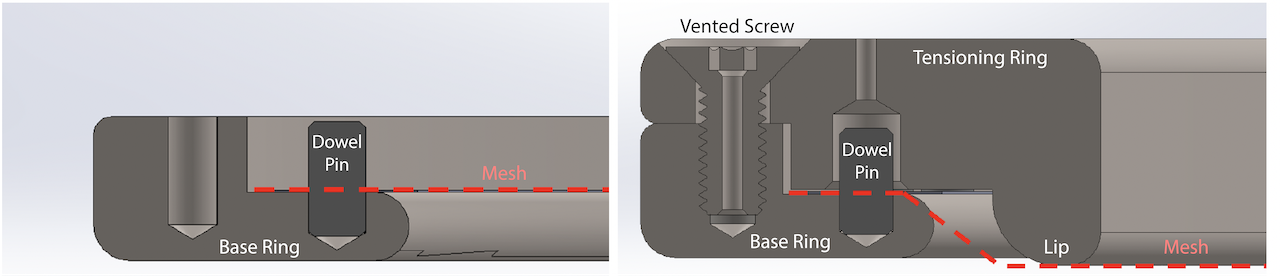}
\caption{\label{fig:NEXT100_frame} Mesh tensioning method used for each EL and cathode frame assembly. Left: dowel pins are inserted into holes around the inside of the base ring and the mesh is laid on top. Right: tensioning ring has been placed on top and bolted closed with vented screws, pressing the mesh downwards and stretching it. Vented screws and through-holes were incorporated into the design to minimize virtual leaks and the ring surface was electropolished to improve the performance under high voltages. }
\end{figure}

Figure \ref{fig:NEXT100_frame} shows a sketch of the tensioning procedure. A total of 180 dowel pins made of phosphor bronze (CuSn8P) are inserted around the base ring which holds the mesh in place. The photoetched mesh contains a 9~mm border with holes of the same diameter and location as the dowel pins (see Fig.~\ref{fig:EL_Mesh}) which are placed on top of the dowel pins. A tensioning ring is then placed on top of the mesh and incrementally screwed down (with 72 CuSn8P screws) until it fully contacts the base ring. The lip of the tensioning frame presses down to slowly stretch the mesh. Feeler gauges are used in the outer gap between the base and tensioning ring to control the uniformity around the ring during each tightening increment. During tensioning, the rings are clamped down to the table to prevent buckling of the thinner base ring until the gap is closed enough to distribute the tension through the vented bolt and frame.  This same design, but on a smaller scale, was used in prototype form for the NEXT-CRAB-0 experiment~\cite{Byrnes_2023}. 

The cathode for NEXT-100 follows the design of an EL frame but is designed for lower tension due to the reduced local field strength ($\sim$2~kV/cm) compared with the EL ($\sim$20~kV/cm).  The tension applied was reduced by decreasing the tensioning lip size. This reduces the amount of frame material required, which in turn offers a small radiopurity advantage. 

Considerations were taken in the design to ensure that the silicon bronze frames would be able to support the mesh tension sufficiently. Silicon bronze has slightly reduced mechanical properties compared with stainless steel (Young's modulus = 200 GPa and 115 GPa and yield strength = 230~MPa and 205 MPa for SS~\cite{MakeItFrom.com_2020} and silicon bronze~\cite{SiliconBz} respectively). The expected mesh tension applied to the frames is estimated with the following formula:

\begin{equation}\label{eqn:meshtensiontheory}
    \frac{3}{2} \times \frac{\Delta L}{L_0} \times YAN,
\end{equation}
where $L_0$ is the diameter of the mesh, $\Delta L$ is the stretch induced by the tensioning frame on the mesh, $Y$ is the Young's modulus, $A$ is the cross sectional area of the wire (sheet width $\times$ wire thickness), and $N = L_0/(\textrm{hexagon diameter}\times\textrm{wire thickness})$. We estimate a max mesh tension of 3.6~kN for the EL frame and 0.6~kN for the cathode frame if fully closed. Tension can be fine-tuned by partially bolting the rings together, providing less stretching to the mesh. 

The critical and estimated value of the section area moment of inertia (buckling point) for the EL and cathode frames were calculated to ensure they would not buckle under a maximum mesh tension of 3.6~kN if the frame is fully closed. We find the estimated maximum load the EL rings can take is more than 5~kN of tension. The distribution of force across the tensioning and base rings prevents buckling of the frame assuming a safety factor of 1.5. Since the cathode is tensioned less than the EL rings it will also be able to hold the tension.

\subsection{EL brackets}\label{sec:elbrackets}
The HDPE brackets that hold the EL frames together are shown in Fig.~\ref{fig:NEXT100_EL_Cathode}. The brackets have a U-shape design with a ridged inner surface to minimize leakage current across the surface of the bracket. The ends of the brackets are used to attach the EL frames directly to the copper end cap of the NEXT-100 detector. The gate frame mounts to the brackets by 16 gate screws made from polyether ether ketone (PEEK).  The 16 anode screws, on the other hand, sit in a low-field region and are made from stainless steel. Measurements of the EL gap distance (before the mesh was tensioned) were taken with vernier calipers to estimate the variation in the gap distance across the frames. The mean and standard deviation were found to be 9.7$\pm$0.15~mm, after accounting for the thickness of two meshes (0.127~mm each). The uncertainty on the gap distance is mostly driven by the manufacturing uniformity of the tensioning ring lip and the bracket surface.

\subsection{Material choice and radioassay}\label{sec:radioassay}
The material choice of each component for the EL and cathode was chosen to meet the radiopurity requirements of the experiment by screening for contamination of $^{238}$U and $^{232}$Th. The target background requirement for NEXT-100 is 1$\times 10^{-3}$ counts/(keV kg year) or $\sim$1 count/yr over the full $0\nu\beta\beta$ energy region of interest (16~keV)~\cite{DiazLopez:2023xxq}. The main radioactive backgrounds in NEXT originate from gamma rays via $^{208}$Tl (2.615 MeV) and $^{214}$Bi (2.448 MeV) which are produced in the natural radioactive chains from $^{232}$Th and $^{238}$U respectively. 

Measurements of the EL and cathode frames were performed using inductively coupled plasma mass spectrometry (ICP-MS) at Pacific Northwest National Laboratory with an Agilent 8900 where quantization was performed using isotope dilution mass spectrometric methods~\cite{LAFERRIERE201593}. The technique is sensitive to trace amounts of $^{232}$Th and $^{238}$U to part-per-trillion (ppt) in a given sample. Estimations of the $^{208}$Tl and $^{214}$Bi contamination are made from the $^{238}$U and $^{232}$Th concentrations assuming secular equilibrium. While the initial frame design was based on stainless steel, we were unable to source plates that could meet the required radiopurity specifications at the necessary scale. For the final parts, a design based on silicon bronze frames was adopted due to the superior radiopurity of available materials.  Previous screening campaigns of various detector components for NEXT-100 are published in Ref.~\cite{VAlvarez_2013}. Samples of stainless steel and silicon bronze plates were taken from various suppliers for testing (with the plate reserved with the manufacturer). The results of this screening campaign are shown in Tab.~\ref{table:radioassay}. The thicker plates correspond to the tensioning ring, while the thinner plates are for the base ring. While there were different vendors for supplying the C65500 grade of silicon bronze, the material certifications indicate the material originated from a single source: \textit{KME Mansfeld GmbH}. For the C64200 grade of silicon bronze, the material certification suggests an origin from \textit{Dongzhou Longhe Metals}. Measurements of silicon bronze C65500 were in single-digit ppt (or just above) levels for all samples. These values were either similar or smaller than the values reported in Ref.~\cite{LEONARD2008490}. The C64200 silicon bronze alloy contained much higher levels. This specific alloy contains aluminum, which has been generally reported to contain large contamination~\cite{LEONARD2008490,Radiopurity}, and has a contamination four orders of magnitude beyond the desired specification for NEXT-100. As a result, the NEXT-100 EL and cathode were chosen to be made from batches 2023-03 and 2023-08 listed in Tab.~\ref{table:radioassay}.

\begin{table}[t]
\begin{center}
\centering
\begin{tabular}{ccccccc}
\hline
Sample  &  Alloy &       Supplier & Thickness &   $^{232}$Th [ppt] &      $^{238}$U [ppt] \\
\hline
& Silicon Bronze \\
2022-17 & C65500 &    \textit{Atlas Metal} &     1/2 in &       4.4$\pm$0.5 &   14.6$\pm$1.3 \\
2022-17 & C65500 &    \textit{Atlas Metal }&     3/4 in &       2.3$\pm$0.3 &    2.3$\pm$0.7 \\
2022-35 & C65500 & \textit{Farmers Copper} &      10 mm &       3.8$\pm$0.5 &    2.2$\pm$0.5 \\
2022-35 & C65500 &\textit{ Farmers Copper} &      15 mm &       3.9$\pm$0.3 &    4.9$\pm$1.1 \\
2023-08 & C65500 & \textit{Farmers Copper} &     1/2 in &       1.4$\pm$0.3 &    4.9$\pm$0.6 \\
2023-08 & C65500 & \textit{Farmers Copper} &     5/8 in &       0.6$\pm$0.4 &    4.7$\pm$0.6 \\
2023-03 & C65500 & \textit{Farmers Copper} &     1/2 in &       0.5$\pm$0.3 &    4.5$\pm$0.3 \\
2023-03 & C65500 & \textit{Farmers Copper} &     5/8 in &      16.3$\pm$0.7 &    5.5$\pm$0.5 \\
2022-27 & C64200 &   \textit{Additive Mfg.} &      10 mm &     25000$\pm$800 & 84000$\pm$9000 \\
2022-27 & C64200 &   \textit{Additive Mfg.} &      15 mm &     34861$\pm$554 & 99797$\pm$3336 \\
\hline
& Stainless Steel \\
2021-37 & SS304 &  \textit{Additive Mfg.} &        10 mm &         104$\pm$8 &     756$\pm$27 \\
2021-37 & SS304 &  \textit{Additive Mfg.} &        15 mm &       3346$\pm$64 &     526$\pm$30 \\
\hline
\end{tabular}
\end{center}
\caption{The various ICP-MS measurements of materials considered for the EL and cathode base and tensioning rings. Each sample measurement was repeated three times, the rows show the combination of these repetitions. The thickness column corresponds to the thickness of the metal plate (which comes in metric or imperial sizes depending on the vendor) that a small 1~cm$^3$ sample is taken from. High levels were detected in the C64200 alloy, a silicon bronze alloy containing aluminum. } \label{table:radioassay}
\end{table}

Measurements of the other components in the EL and cathode region are shown in Tab.~\ref{table:radioassay2}. Measurements were performed using either ICP-MS, Glow Discharge Mass Spectrometry (GDMS) with \textit{Eurofins}~\cite{eurofins}, or with High Purity Germanium (HPGe) detectors at the Sanford Underground Research Facility~\cite{akerib2020luxzeplin}. 

\begin{table}[t]
\begin{center}
\centering
\begin{tabular}{cccccc}
\hline
Component  &  Material &  Method &   $^{232}$Th [ppt] &      $^{238}$U [ppt] & Total Mass\\
\hline
EL/cathode mesh         &  SS316Ti &  ICP-MS &  202.3$\pm$6.8   &    604.1$\pm$38.4 &  509 g   \\
Dowel pins and screws   &  CuSn8P  &  GDMS   &  $<$100          &    $<$100         &  329 g   \\
EL bracket              &  HDPE    &  ICP-MS &  15$\pm$5        &    3$\pm$1        &  324 g   \\
EL bracket gate screws  &  PEEK    &  HPGe   &  $<$4.1         &    $<$0.6 &  2.7 g   \\
EL bracket anode screws &  18-8 SS &  HPGe   &  2630$\pm$1490   &    480$\pm$230    &  22.4 g   \\
\hline
\end{tabular}
\end{center}
\caption{The measurements and total mass of the other components in the EL and cathode regions. There are 540 dowel pins and 216 CuSn8P screws.} \label{table:radioassay2}
\end{table}

The EL and cathode frames contribute the vast majority of the mass of these components (26.1~kg). A contamination level of <10 ppt is sufficient for meeting the target requirements for the frames, while looser requirements are implied for the mesh, screws, and dowel pins due to their lower mass.  The total expected contribution to the background index of NEXT-100 from the EL regions and cathodes as assayed is 7.4$\times10^{-3}$~counts/yr and 8.8$\times10^{-2}$~counts/yr for $^{208}$Tl and $^{214}$Bi respectively, meeting the required specification.

\section{Electrostatic breakdown of the EL region}\label{sec:breakdown}

\subsection{Mesh robustness under discharges}
The NEXT-100 EL region has a minimum target electric field specification of 15~kV/cm at 13.5~bar such that the electrons can acquire sufficient energy to excite the xenon medium to produce scintillation light and the EL gain is non-limiting for fulfilling <1\% energy resolution. The option to run at higher voltages to achieve larger gains is also desirable, up to a maximum anticipated operating voltage around 28~kV, just below the transition to avalanche.  To achieve this, the mesh must have a high-quality surface finish to avoid breakdowns and be robust in the event of the inevitable occasional sparks to avoid permanent damage from discharges.

\begin{figure}[b]
\centering
\includegraphics[width=0.31\linewidth]{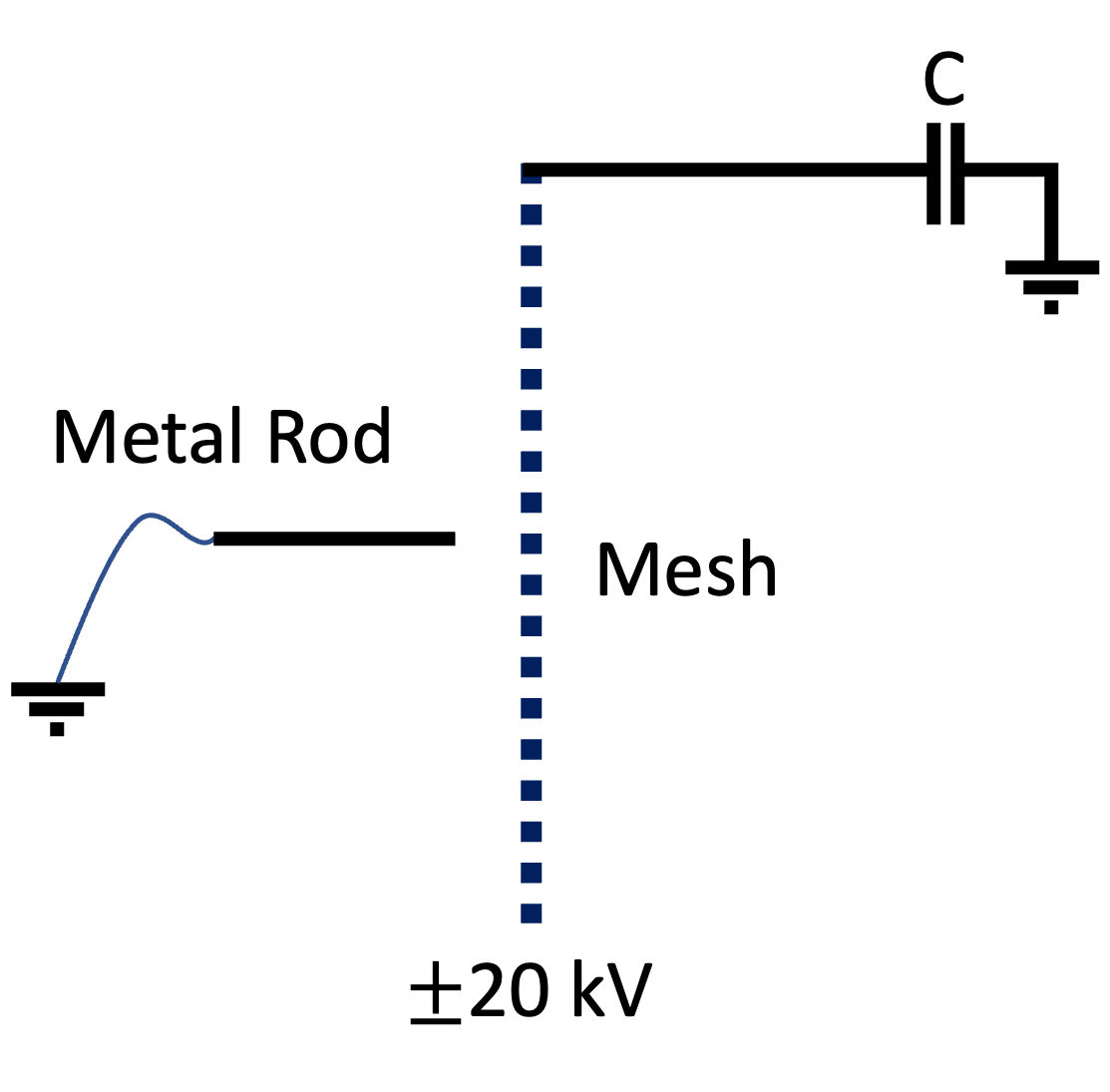}
\includegraphics[width=0.25\linewidth]{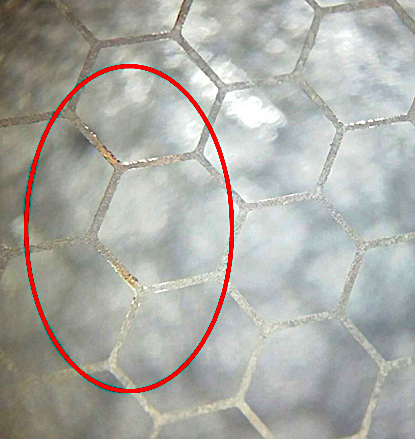}
\includegraphics[width=0.29\linewidth]{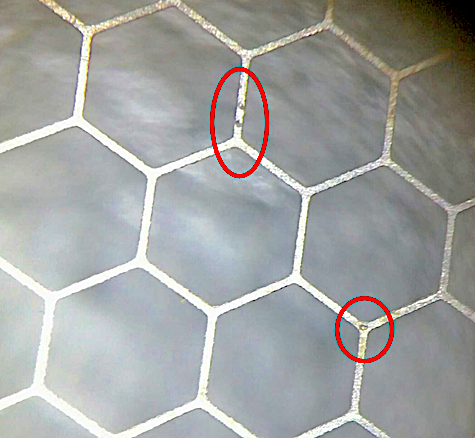} 
\caption{\label{fig:sparkrobustness} Left: setup for high voltage sparks to the photoetched mesh. Middle: picture of sparks with a 10~nF capacitor and $-20$~kV. Some shiny areas could be seen but no breaks or surface damage. Right: picture of sparks with a 10~nF capacitor and $+20$~kV. Some minor surface damage was visible under the microscope but no breaks or holes through the mesh were found.}
\end{figure}

Electrically the EL region can be approximately modeled as a large parallel plate capacitor with xenon gas as the dielectric medium. Considering an operating voltage of 28~kV, the meshes at the radius of NEXT-100 would have a capacitance of 1.4~nF and discharge with 0.52~J of energy. This is a factor of around eight larger than the energy stored in the smaller NEXT-White EL region at the same voltage~\cite{NEXT:2018rgj}, giving the potential for significantly more damaging discharges and a need for substantially higher material robustness. 

To test the electrical robustness of the meshes under discharge, a capacitor was attached to a mesh sample held at voltage, as shown in Fig.~\ref{fig:sparkrobustness}. A thin, grounded metal rod was moved toward the mesh until a discharge occurred. The test was carried out in air and repeated for both positive and negative polarities at voltages up to 20~kV, and capacitances of 1, 5, and 10~nF. For discharges at energies up to 2~J no structural damage was found. At 2~J, (about a factor of 4 times the maximum operating conditions), some barely visible surface features were observed, with no apparent structural damage.  With the mesh held at negative voltage, the surface became shinier where some surface material was ablated, whereas at positive voltage some small-scale pock marks were observed at the discharge points. These results demonstrate that photoetched meshes of this thickness are robust in the electrical discharge conditions of the EL region. This is in contrast to ultra-thin woven wire meshes of past NEXT detectors, where discharges of this magnitude can lead to melting and wire breakage.

\subsection{Profiling of breakdown locations}

Preliminary iterations of meshes for the NEXT-100 experiment were defective upon manufacture and contained several broken hexagons on each mesh. Studies of the electric breakdown potential across the EL gap were carried out using these meshes to study the impact of imperfections/broken wire lands. These meshes were stretched in  stainless steel frames identical to the NEXT-100 EL design. The breakdown across the meshes was studied in air at gap distances of 1.1 cm (slightly different from the gap used in NEXT-100). Tests were made by connecting one EL frame to high voltage and the other frame at ground, similar to the operation in NEXT-100. High voltage is supplied via a GLASSMAN PS/EG30N1-100TT 30 kV high-voltage supply, operated up to 20 kV. In each test, the voltage was slowly ramped up until a spark was initiated across the gap from the mesh surface. The supply was ramped back down to zero and switched off to reset the system after each spark. 

\begin{figure}[t]
\centering
\includegraphics[width=0.4\textwidth]{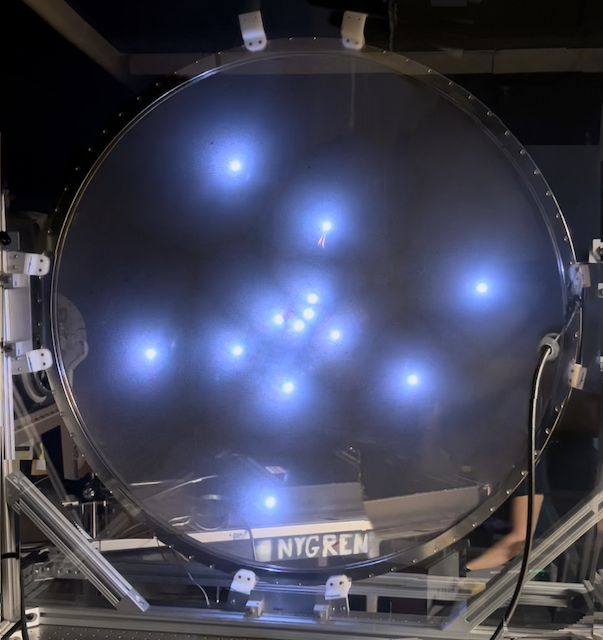}
\includegraphics[width=0.48\textwidth]{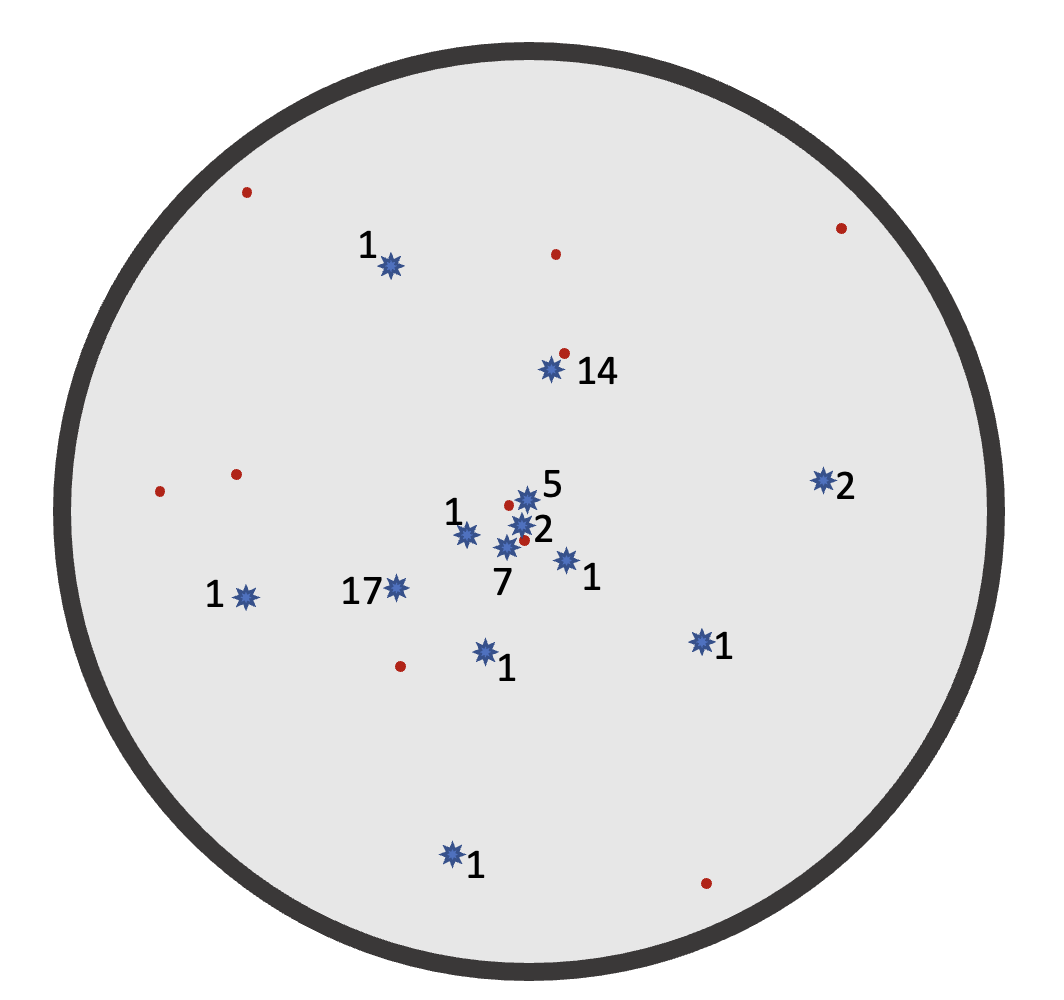}
\caption{\label{fig:spark} Left: composite image of the sparks on the EL region. Right: a diagram of the sparks with positions of breaks/defects shown by the red dots against the sparks shown by the blue stars. The numbers indicate the number of sparks at a spot. }
\end{figure}

Figure~\ref{fig:spark} shows the results of the tests. The red dots indicate positions in the mesh where a defect (either a break on the wire land, a sharp edge, or some other visible surface variation) was identified while the numbers by the blue stars indicate the number of sparks at a given position. Sparks occurred with a higher frequency towards the center of the meshes where there is the largest expected electrostatic deflection (see Sec.~\ref{sec:deflection}). Some sparks occurred with a position close to a defect, though there was no evidence of a strong correlation between spark points and defect points. The breakdown potential consistently ranged from 17.5 -- 18.5 kV, roughly the expected breakdown field strength for air at 1~bar (22 kV)~\cite{ALATOUM2020107357}. Notably, this is already a sufficient voltage to run NEXT-100 with reasonable gain for physics analyses in xenon at 13.5~bar, achieved even in the much weaker dielectric medium of air at one atmosphere. 

\section{Electrostatic deflection}\label{sec:deflection}

Due to the high voltage in which the EL region operates, the electrostatic force between the two parallel meshes can be significant enough to pull the mesh surfaces towards one another causing a non-planar surface. This can cause a non-uniform electric field leading to a non-uniform gain and increased chance of sparking. In this section, we provide analytical calculations and measurements of the NEXT-100 EL deflection under high voltage in various configurations. The impact of the deflection under different operating conditions is also discussed. 

\subsection{Analytical calculations}
\label{sec:predictions}
A derivation of the expected electrostatic deflection is given in Appendix~\ref{appdix:deflection}. In summary, the electrostatic deflection, $z$, of a symmetrical circular mesh is given by the equation:
\begin{equation}
\label{eq:deflection}
z = -\kappa(R^2 - \rho^2),\quad \kappa = \frac{\epsilon RE^2}{4\tau}
\end{equation}
where $\epsilon$ is the permittivity, $R$ is the radius of the mesh, $\rho$ is the radial distance from the center of the mesh, $\tau$ is the tension of the mesh, and $E$ is the electric field between the meshes with electrical potential difference, $V$, separated by a gap $g$, $E = V/g$. The maximum deflection occurs in the center of the mesh at $\rho = 0$ cm,
\begin{equation}
\label{eq:max_deflection}
z^{\textrm{max}} = -\kappa R^2.
\end{equation}
We find that the deflection does not depend on the Young's modulus of the material and has a quadratic dependence on the electric field, $E$. Properties of the mesh such as the thickness and hexagon size factor into the tension of the mesh.

\begin{figure}[t]
\centering
\includegraphics[width=0.65\textwidth]{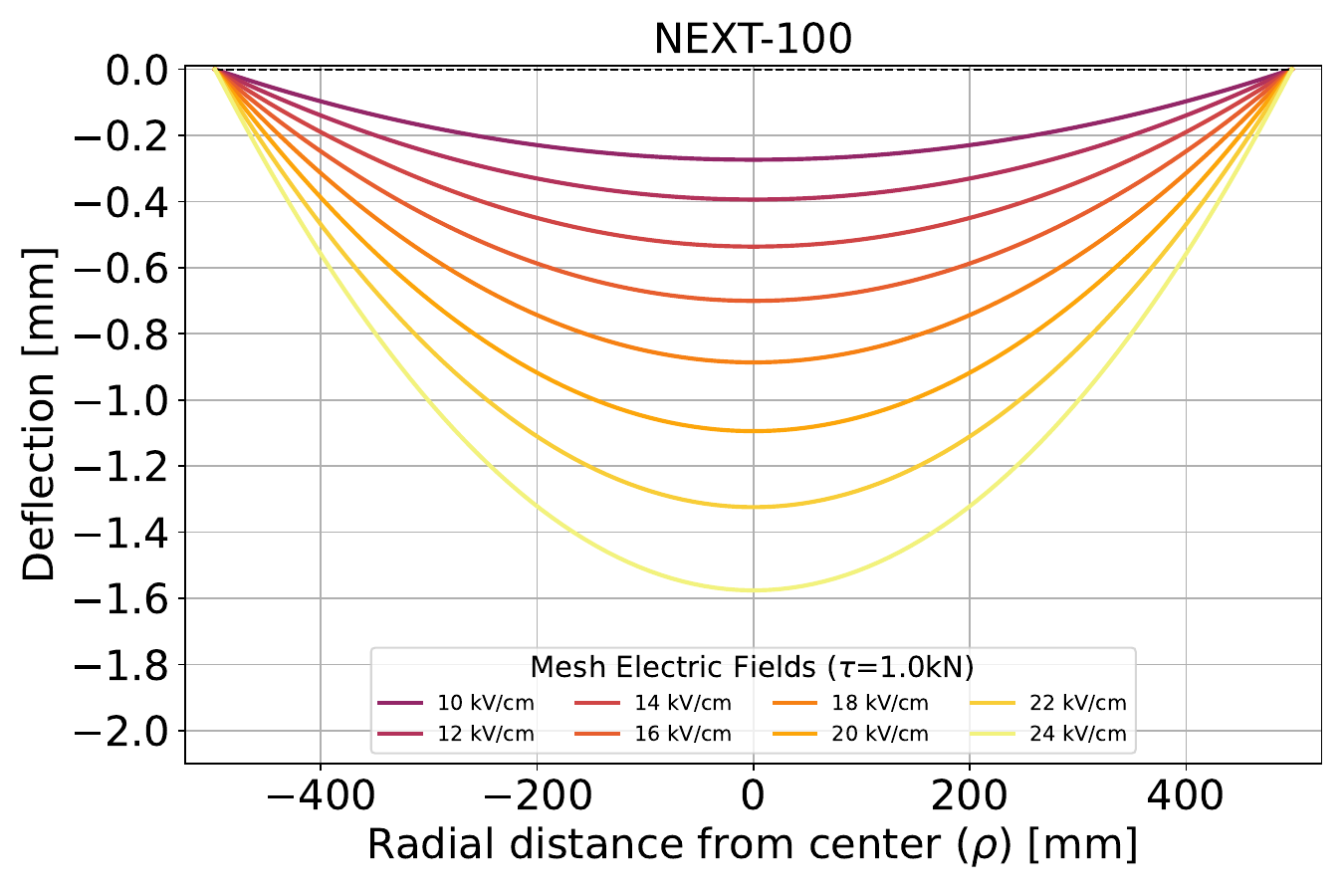}
\caption{\label{fig:pred_deflecion} The predicted electrostatic deflection for a NEXT-100 mesh with a tension of 1~kN. The various lines show the deflection at different electric fields. This deflection reduces the EL gap distance of NEXT-100 (9.7$\pm$0.15~mm).}
\end{figure}

Figure~\ref{fig:pred_deflecion} shows the predicted electrostatic deflection for a NEXT-100-sized mesh at 1.02~m at several electric field values for a fixed mesh tension of 1~kN. With each of the EL gate and anode meshes deflecting towards each other, the total deflection is given by twice the values shown in Fig.~\ref{fig:pred_deflecion}. 

Predictions and measurements with a support post of the type studied in Ref.~\cite{NEXT:2018wtg} placed between the meshes to reduce deflection are given in Appendix~\ref{appdix:deflection} and \ref{appdix:deflection_measurements} respectively. It was decided against including the support post in the NEXT-100 experiment due to the satisfactory performance of an unsupported mesh. 

\subsection{Measurements of electrostatic deflection}
\label{sec:results}

\begin{figure}[t]
\centering
\includegraphics[width=\textwidth]{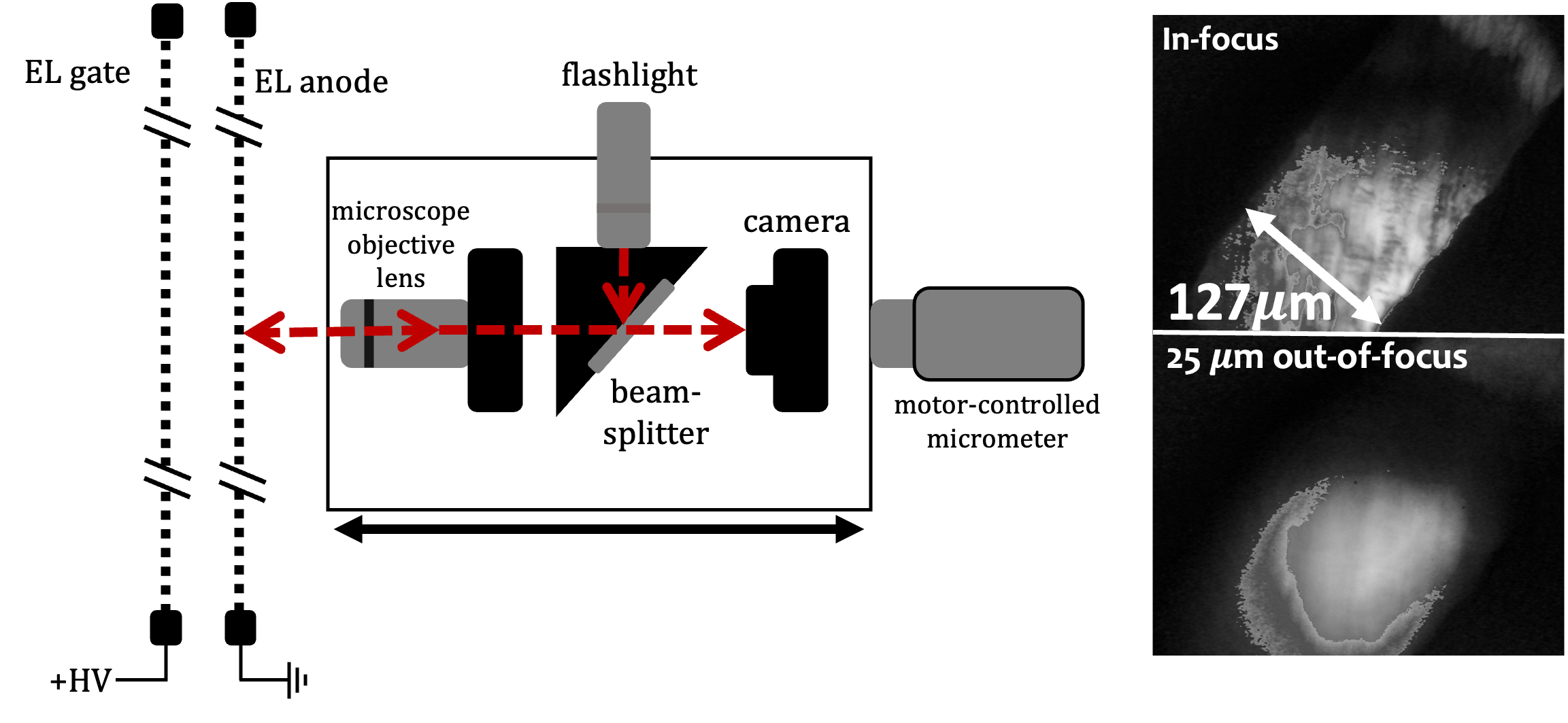}
\caption{\label{fig:deflection_optics} Left: a schematic of the optical system used to measure the electrostatic deflection of the mesh. Right: an image of an in and out-of-focus mesh as viewed from the webcam. The EL gate and anode are held together using insulating brackets made from HDPE.  }
\end{figure}

The electrostatic deflection was measured using the apparatus shown in Fig.~\ref{fig:deflection_optics}. The deflection measurement involves moving a micrometer stage containing optical apparatus to find the relative change in focus of the mesh after a voltage is applied. This technique is similar to an approach outlined in Ref.~\cite{LINEHAN2022165955} that we became aware of during the final preparation of this manuscript. 

Light from an LED is reflected off a 1-inch Thorlabs EBS1 50:50 beam-splitter through an Olympus Plan N microscope objective lens (factor 20x magnification) onto the mesh surface. The reflected light is imaged with a Logitech C615 HD webcam to give a real-time image of the mesh. In order to bring the mesh into focus via the objective lens alone, the main webcam lens was removed. The apparatus is mounted onto a movable micrometer stage whose displacement from the mesh is adjusted using a Thorlabs ZST213 actuator controlled by a brushed DC servomotor (Thorlabs KDC101) to determine the distance deflected relative to no potential applied. The actuator has a precision of 5~\si{\micro\meter}. The position of the apparatus is adjustable to measure the deflection across various radii on the mesh surface. A bias voltage is applied to the gate mesh (with the anode at ground) using a GLASSMAN PS/EG30N1-100TT 30~kV high-voltage supply and was varied up to a voltage before breakdown could occur between the meshes. Between each measurement, the voltage was set to 0 V to allow the mesh to reset back to the origin and then slowly ramped up. The focus was also moved far from the focal point between each measurement.

Several datasets of the electrostatic deflection were taken using an identical EL region in design as the EL region to be used in NEXT-100\footnote{This part was procured for installation in NEXT-100 but failed to meet radiopurity specifications (see stainless steel measurements in Tab.~\ref{table:radioassay}).}. Among these tests, studies were made of the deflection as a function of the frame closure during tensioning. Measurements were taken using a fine voltage and radius increment to understand the tensioning protocol in detail. A less granular set of measurements was made on the NEXT-100 region in situ immediately before installation in the detector, to validate successful construction and tensioning. 

\begin{figure}[t]
\centering
\includegraphics[width=0.6\textwidth]{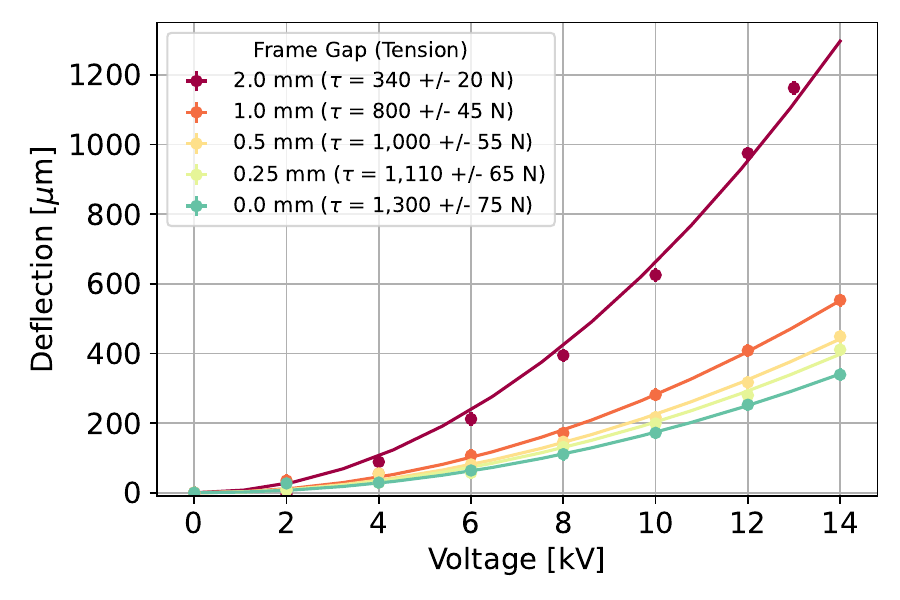}
\caption{\label{fig:def_tensions} The electrostatic deflection at the center as a function of different mesh tensions. The tension values are obtained from a fit (shown by the solid lines) to the data using Eqn.~\ref{eq:deflection}.}
\end{figure}

Fits to these data are performed using Eqn.~\ref{eq:deflection} leaving the tension as a free parameter and minimizing a chi-square metric. Each data point has a deflection uncertainty of 8~\si{\micro\meter} from determining the focus of the mesh, established by taking several repeated measurements at the same position. We estimate the uncertainty from the repeatability of $x$ and $y$ positioning by taking the standard deviation of measurements from repeated placement at the same location (typically ranged from 13 -- 17~\si{\micro\meter}) which is combined in quadrature with the 8~\si{\micro\meter} focusing uncertainty. The fit varies the EL gap distance (1.1$\pm$0.025 cm) and the voltage values (uncertainty of 100~V) randomly sampling 1000 times within their uncertainties.  The fit results for the extracted tension (mean and standard deviation from the sampling) are shown in the legend of Fig.~\ref{fig:def_tensions}.

Figure~\ref{fig:def_tensions} shows the results of the maximum deflection during prototype mesh tensioning.  For a loosely tensioned mesh (with a 2~mm gap between the rings), the deflection reaches 1.2~mm at 14~kV. At full tension, the deflection is reduced to $\sim$0.3 mm, almost a 1~mm total reduction for each mesh. We find that the reduction in deflection from a frame gap of 1~mm compared to fully closed was marginal ($\sim$0.2 mm, see Fig.~\ref{fig:def_tensions}), so it was decided to tension the NEXT-100 meshes to be installed with a 1~mm frame gap. This choice minimizes the chance of breaking the mesh from overstretching. Interestingly, the measurement of the fully tensioned frames is about a factor of 4 lower than the estimations detailed in Sec.~\ref{sec:elframe}. While this leads to a deflection that is slightly more than anticipated, we show in Sec.~\ref{sec:resolution} that the performance for NEXT-100 is suitable. Possible explanations could be that the wires at the local region around the edge of the frame are getting stretched more than the central region and entering plastic rather than elastic deformation, or that the mesh does not slip sufficiently well enough across the tensioning ring surface.

\begin{figure}[t]
\centering
\includegraphics[width=0.48\textwidth]{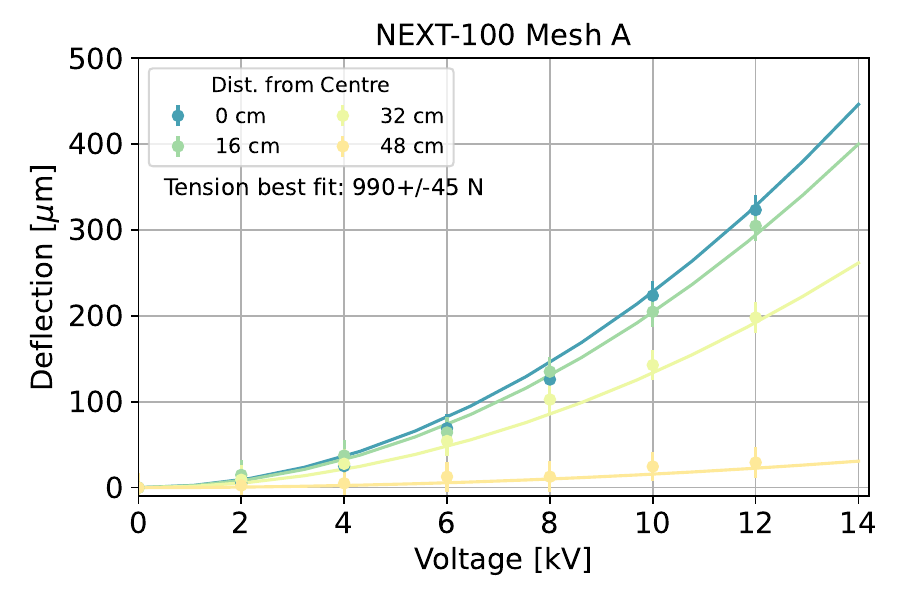}
\includegraphics[width=0.48\textwidth]{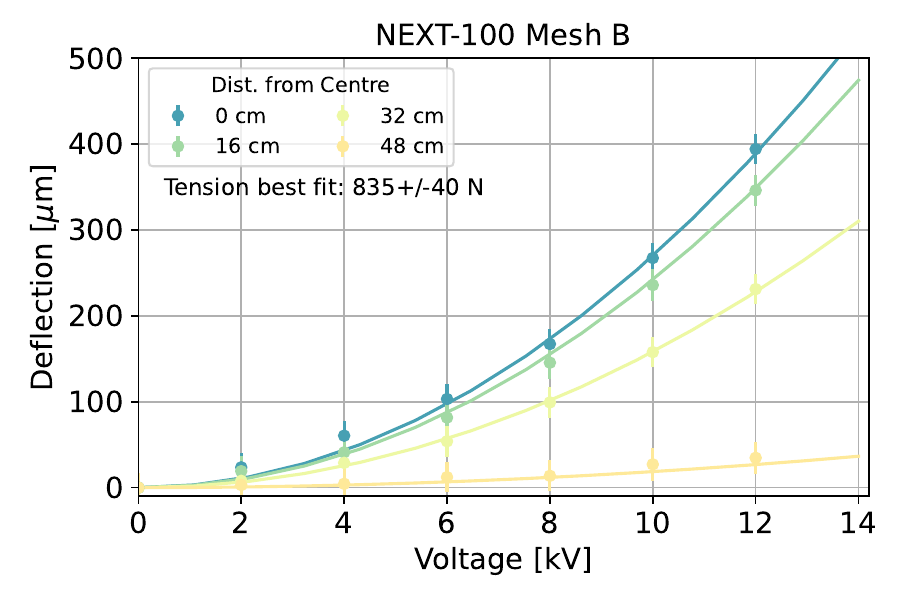}
\caption{\label{fig:sibr_def} The electrostatic deflection for the NEXT-100 mesh A and B. The tension values are obtained from a fit (shown by the solid lines) to the data using Eqn.~\ref{eq:max_deflection}.  }
\end{figure}

Figure~\ref{fig:sibr_def} shows the electrostatic deflection measured on each of the NEXT-100 EL meshes (labeled A and B), as prepared for installation in the Laboratorio Subterr\'aneo de Canfranc (LSC) laboratory. Measurements were taken in 21 different places on the mesh surface covering four radii of 0~cm, 16~cm, 32~cm, and 48~cm with 1, 8, 7, and 5 measurements at each spot respectively. Some locations were not reachable with the apparatus resulting in differences in the number of repeated measurements at a given radius. We find the tension of mesh A to be 990$\pm$45 N while the tension of mesh B to be slightly less at 835$\pm$40 N due to its slightly larger measured deflection. The measurements of mesh B are consistent with the expected 1~mm frame gap tension shown in Fig.~\ref{fig:def_tensions}, however, the tension in mesh A is more consistent with a frame gap of 0.5~mm. On the basis of these data, the maximal expected deflection of each mesh at the 13.5~bar EL threshold is expected to range from 0.35 -- 0.4~mm for mesh A and B respectively.

\subsection{Energy resolution }
\label{sec:resolution}

\begin{figure}[t]
\centering
 \includegraphics[width=0.48\textwidth]{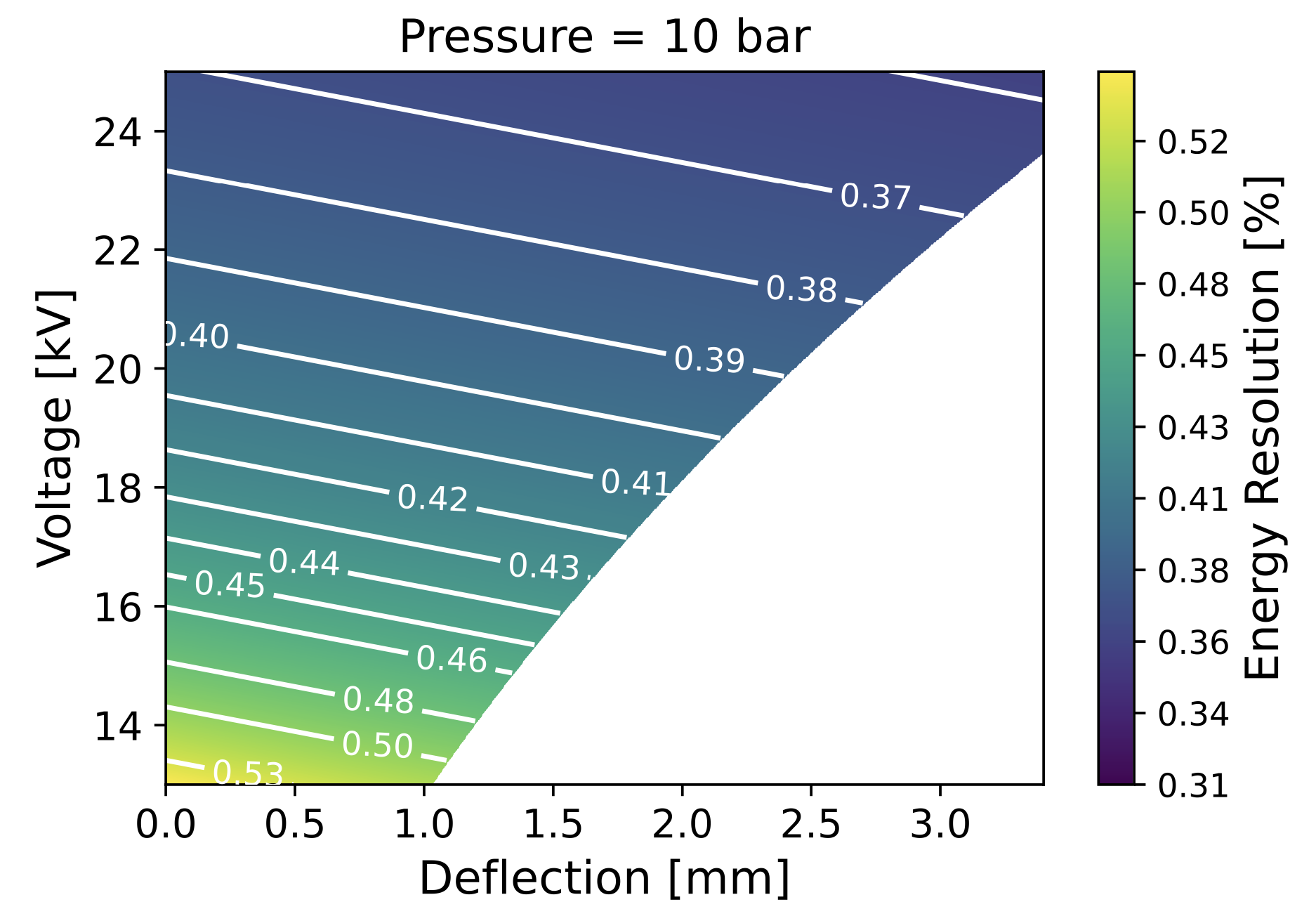}
\includegraphics[width=0.48\textwidth]{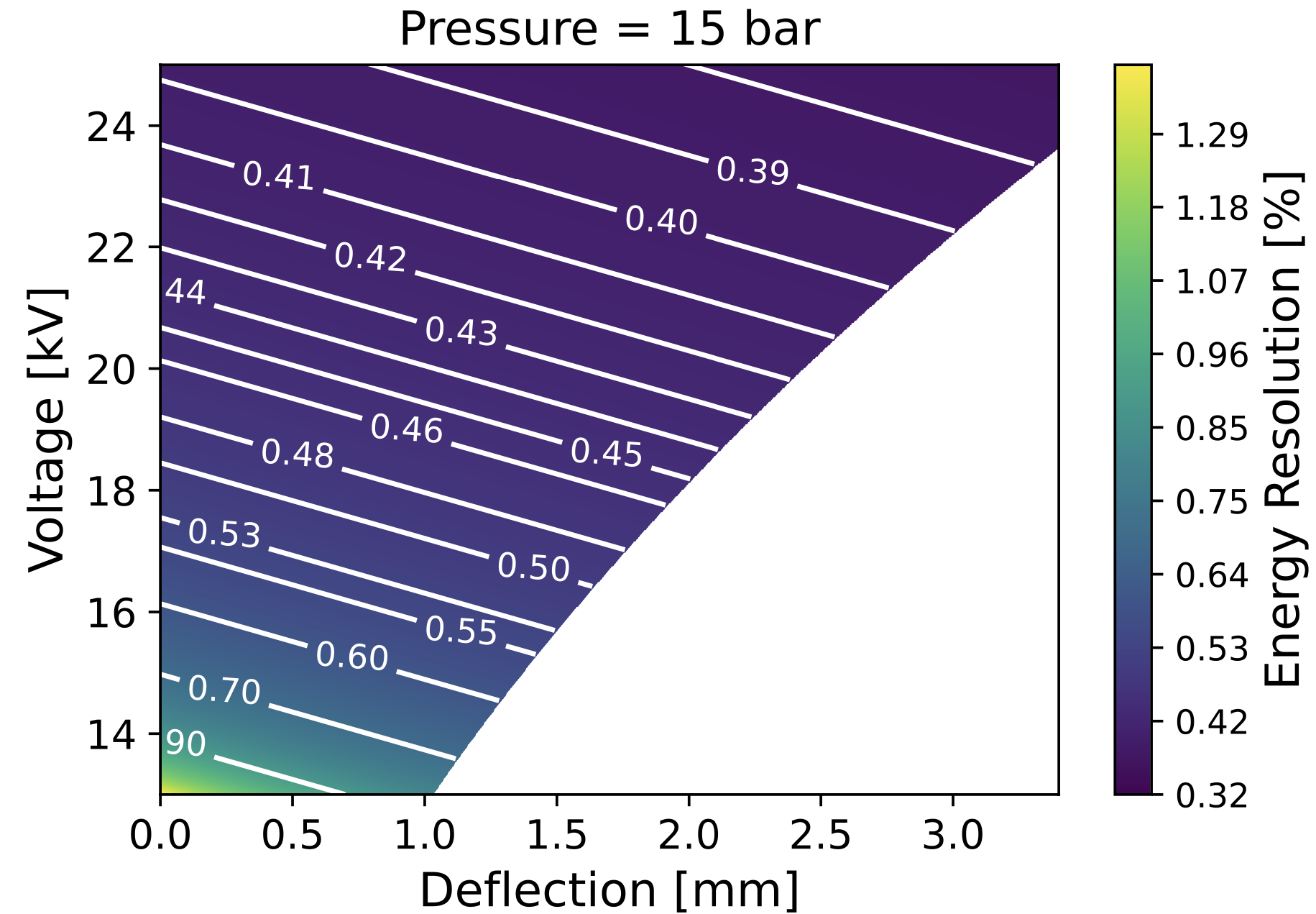}
\caption{\label{fig:eres_pressure} The dependence of energy resolution on deflection and voltage at 10~bar (left) and 15~bar (right) using a mesh tension of 900~N. A factor of two is included in the maximum deflection to account for two deflecting meshes.  }
\end{figure}

Deflection of the meshes indirectly impacts the energy resolution of a $0\nu\beta\beta$ decay event at 2.5~MeV from variation in the gap distance across the surface, altering the EL gain uniformity. To first order this effect is vanishing, since nearly all of the potential energy of the thermalized ionization electrons crossing the gap is converted into excitations that lead to photons, and the total potential energy is independent of the gap size. However, the threshold behavior of the EL process slightly breaks the correspondence between potential and light yield, and some residual gap size dependence is present. We assume no loss in signal due to a finite electron lifetime and no contributions from electronics noise. Both of these effects are expected to be sub-dominant in the measurement of energy resolution and can be controlled for via calibration with $^{83\textnormal{m}}$Kr~\cite{martinez2018calibration}.  We calculate the expected variation in energy resolution considering the fluctuations in the ionization electrons produced (described in Ref.~\cite{NYGREN2009337}), and the change in the gain across the surface due to the gap length and electric field strength using the empirical data provided in Ref.~\cite{CMB_Monteiro_2007}. A uniform electric field at each gap position is also assumed. The results as a function of the electrostatic deflection and bias voltage for a gas pressure of 10 and 15 bar are shown in Fig.~\ref{fig:eres_pressure}. The edge of the white region corresponds to the maximum electrostatic deflection expected for a mesh (Eqn.~\ref{eq:max_deflection}) of the same radius as NEXT-100 and at a tension of $\sim$900 N, similar to the measured values obtained. The deflection values are multiplied by two to account for two deflecting meshes. At each voltage, the deflection will vary from 0 mm (mesh edge) to the max deflection (mesh center) and the energy resolution varies between these extremes. Overall the resolution is below 1\% in almost all areas meeting the desired threshold for operation and variations do not significantly span more than 0.2\% across the mesh surface, although most cases are much below this. We find that at higher voltages the variation in gain due to the deflection is not significant and is compensated for by the increase in overall gain from the higher field.   

\section{Simulation of a small-scale EL and cathode region}\label{sec:simulations}

Uniformity of EL region response is crucial for achieving precise energy measurements through EL amplification. Local non-uniformities in the electric field on the scale of the mesh pattern depend on the relative position and alignment of the two hexagonal meshes, which impacts both the average and the spread of gains across the EL region.  To study these effects a suite of simulations was performed. We use {\tt SolidWorks}~\cite{solidworks} to create a geometrical model of a smaller-scale EL and cathode region of several diameters, thicknesses, and alignments. From the hexagonal symmetry, the small-scale model can be extrapolated to the full-scale. We import these models into {\tt COMSOL Multiphysics} which calculates the the electric field distribution in three dimensions. A {\tt Garfield++} simulation is then used with the {\tt COMSOL} field map to simulate individual electron trajectories and study the impact on EL gain and how this factors into energy resolution. 
\subsection{EL electric field uniformity}\label{sec:comsol}
\begin{figure}[t]
\centering
\includegraphics[width=0.51\textwidth]{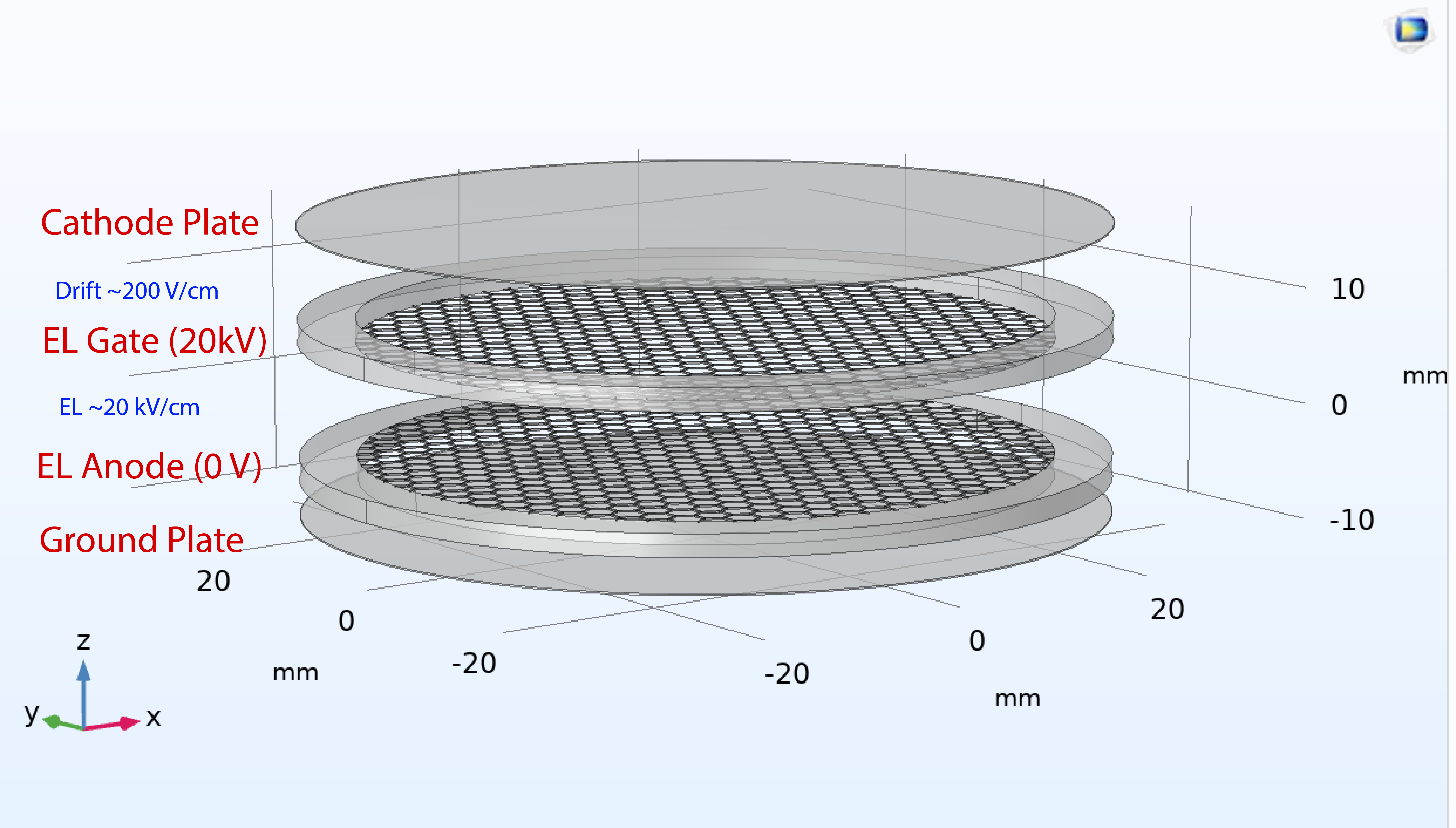}
\includegraphics[width=0.46\textwidth]{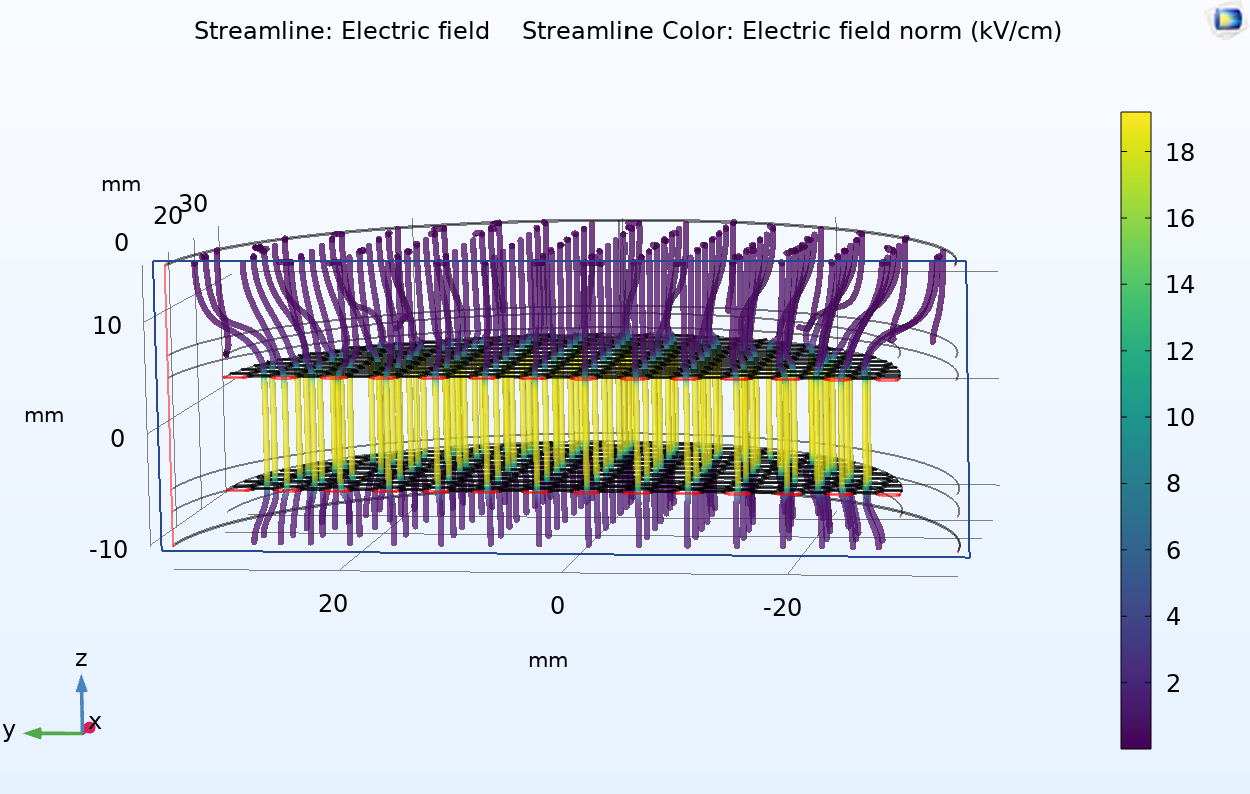}
\caption{\label{fig:COMSOL} Left: the simplified model of the small EL region in {\tt COMSOL} with a 60 mm diameter mesh. Right: the electric field lines viewed in a slice of the $x,y$ plane simulated with {\tt COMSOL}. Due to the shape of the hexagons, the field lines are funneled into the center of the hexagons. The electric field changes from a low drift field of 200~V/cm to a high field region of 18.8~kV/cm at the EL gate mesh surface. The meshes are separated with a 1 cm gap distance. }
\end{figure}

Figure~\ref{fig:COMSOL} shows the electric field model of an EL region with each frame of thickness 1~cm consisting of a mesh diameter of 60~mm at the surface. The hexagonal geometry of the mesh is similar to the NEXT-100 EL region design with 2.5~mm inner diameter hexagons and 127~\si{\micro\meter} thickness and wire width. The EL frames are spaced at a 1~cm gap distance and 20~kV is applied to the gate frame. We include a grounded plate below the meshes to terminate the field lines and a cathode plate to set a drift field $\sim$200 V/cm above the EL gate frame. This models the field lines as they would originate from the drift region and terminate at a grounded tracking plane sitting closely behind the anode. As the field lines cross the gate mesh boundary, they increase in value to around 20 kV/cm before decreasing to 0 kV/cm as they exit and terminate at the grounding plate.

\begin{figure}[t]
\centering
\includegraphics[width=0.5\textwidth]{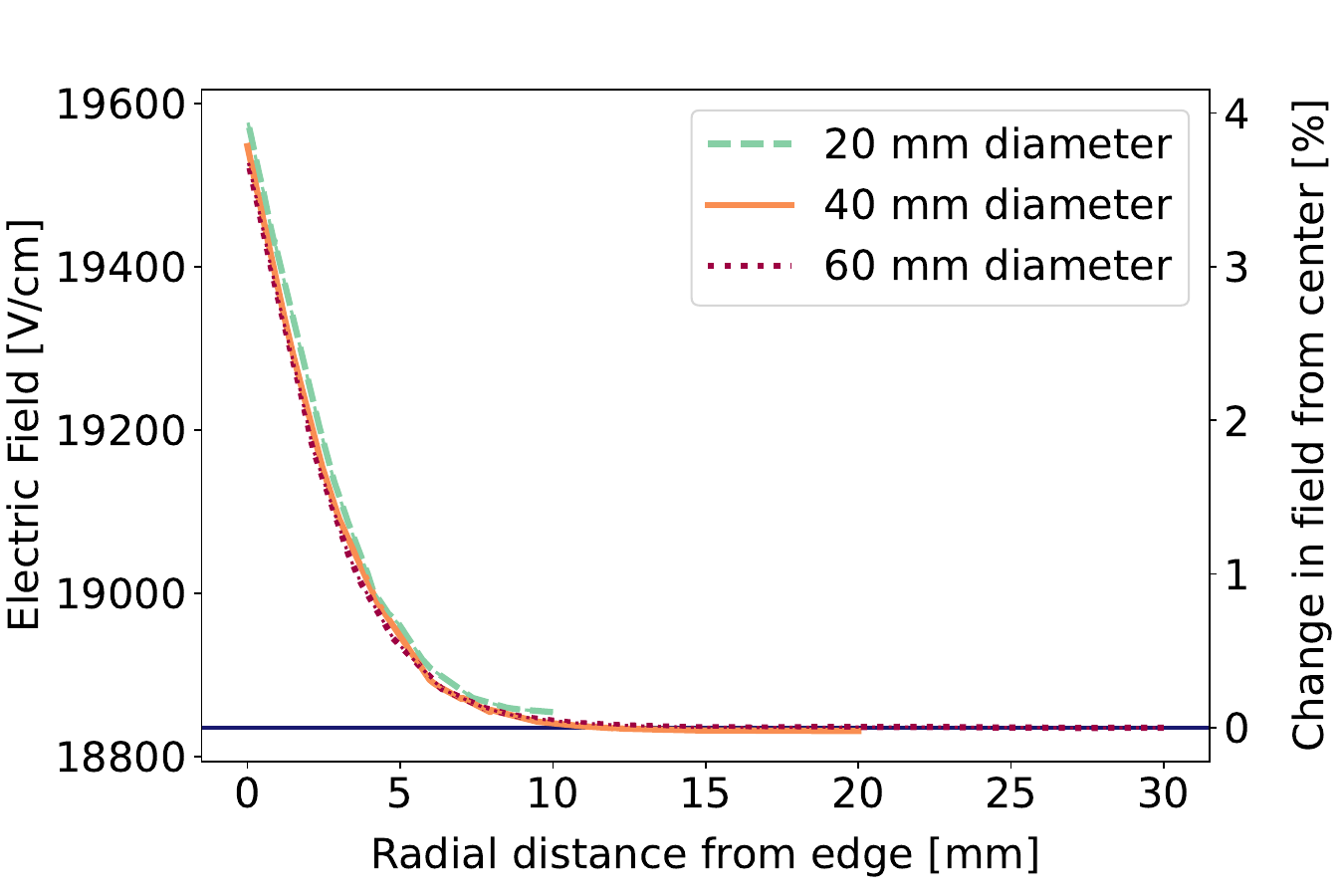}
\raisebox{0.35\height}{\includegraphics[width=0.46\textwidth]{plots/COMSOL_SideView.png}}
\caption{\label{fig:E_field_radius} Left: the electric field from the edge of the EL region radially inwards to the center for different mesh diameters. The blue line shows the electric field value at the center of the 60~mm EL region. Right: a side view of the 60~mm diameter model, the red arrow depicts the direction of the radial field studied. }
\end{figure}

The electric field starting from the edge of the EL region going radially inwards is studied for an EL region with different diameters of 20~mm, 40~mm, and 60~mm. The electric field for each of these configurations is shown in Fig.~\ref{fig:E_field_radius}. We find that the electric field in the center is slightly reduced from the ideal parallel plate scenario of 20 kV/cm with an average field closer to 18.8~kV/cm, roughly a 6\% reduction. Additionally, there is approximately an 11~mm region from the inner edge of the frame where the electric field is not uniform. In these geometries, the size of this non-uniform region does not depend much on the mesh size. The non-uniform region is due to the field transitioning to the parallel plate scenario (with a field of 20~kV/cm) where the EL frames lie which are two parallel metal surfaces. 

\subsection{Cathode field leakage}\label{sec:cathodefringe}
The variation in the electric field from the buffer region to the drift region was investigated to inform the mesh parameters for the cathode. Since the cathode is highly transparent and the buffer electric field is much stronger than the drift field (due to design constraints), there will be a small region beyond the cathode mesh surface where the buffer electric field penetrates into the drift region (field leakage). Very near the cathode, ionization electrons produced near the cathode surface may travel to the buffer region rather than the EL region causing a loss in efficiency.  The field leakage is non-uniform and depends on the distance to the nearest hexagon center. We calculated the maximum depth reached by the fringe field, which occurs at the center of each hexagon. 

\begin{figure}[t!]
\centering
\includegraphics[width=0.48\linewidth]{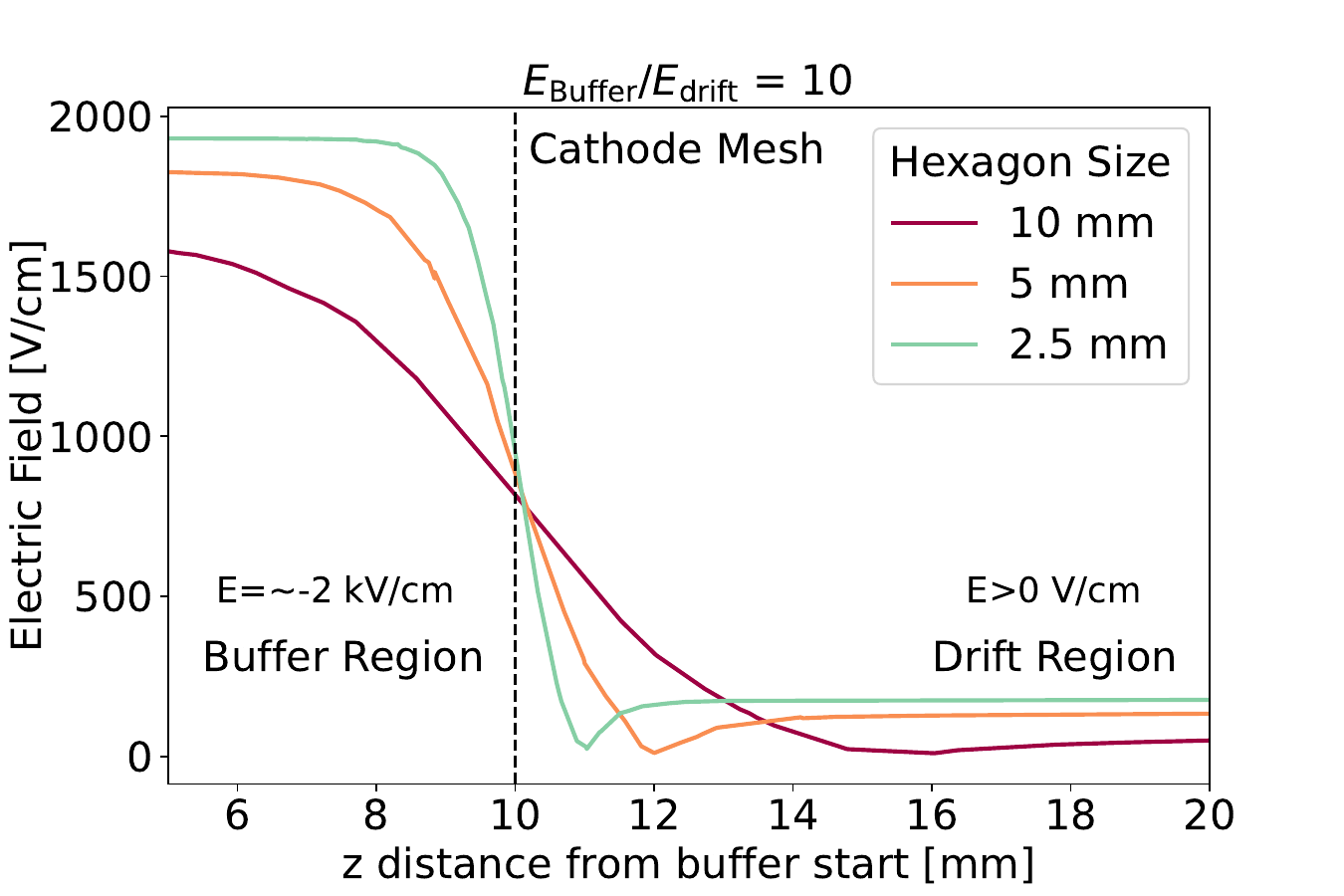}
\includegraphics[width=0.48\linewidth]{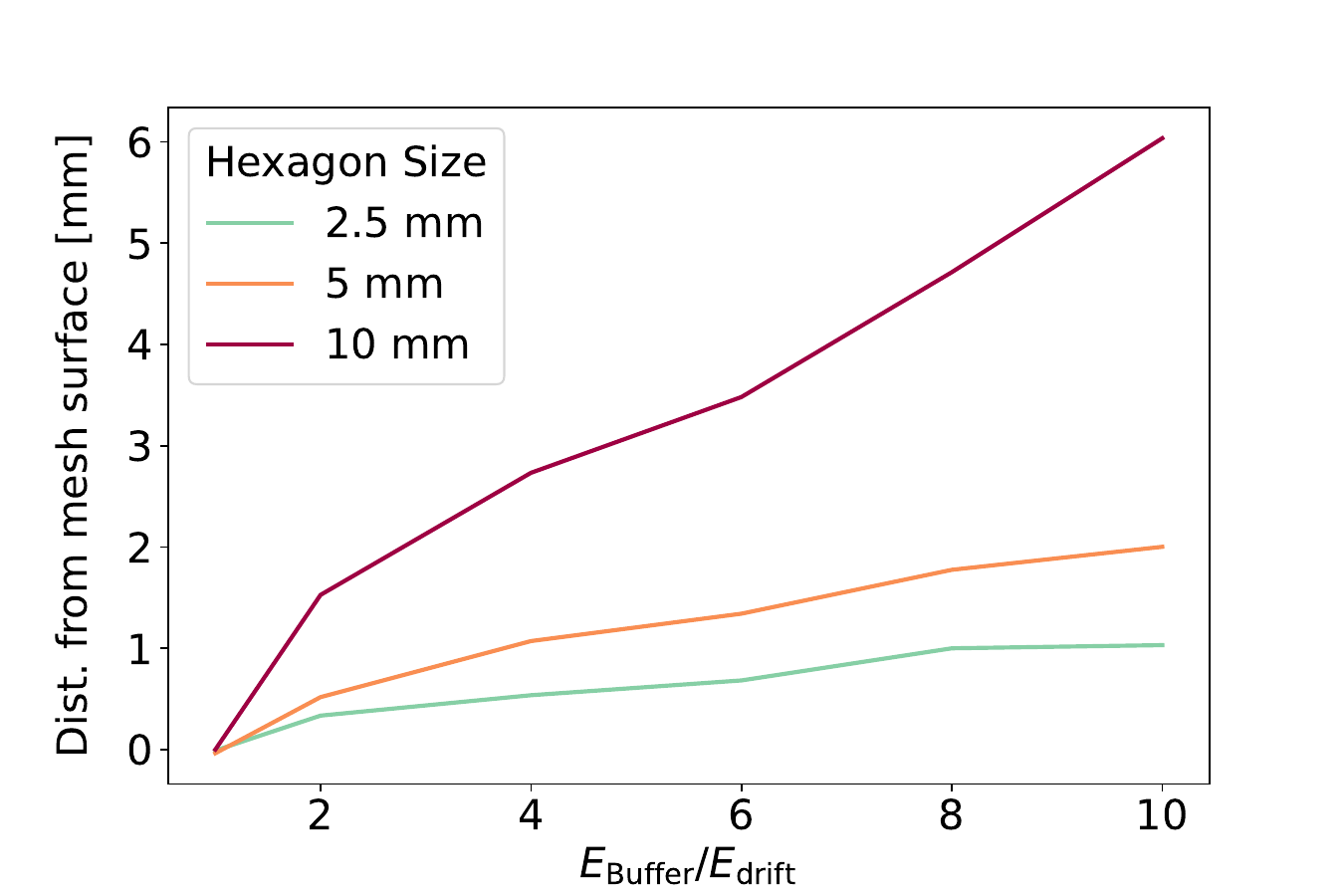}
\caption{Left: the electric field from the buffer region to the drift region as a function of different hexagon sizes. The minima of each curve correspond to the maximum fringe field depth given at the center of the hexagon. Right: the maximum fringe field depth (distance from $z$=10~mm into drift region to the electric field minimum) as a function of different buffer to drift field values at each hexagon size. The expected NEXT-100 $E_{\mathrm{buffer}}$/$E_{\mathrm{drift}}$ ratio is 7 which has a similar fringe field depth for the 2.5 and 5~mm hexagons.}
\label{fig:fringefields}
\end{figure}

Several hexagon sizes (2.5~mm, 5~mm, 10~mm inner diameter) for the cathode mesh were investigated at different buffer-to-drift field ratios. Two disks were placed on either side of a 60~mm mesh diameter cathode frame in {\tt COMSOL} to fix the field boundary conditions. An electric field of $\sim$2~kV/cm was used in the buffer region,  and the drift field was varied to different values. The position of the minimum of the electric field from the cathode mesh surface is determined to estimate the depth of the leakage field. The results of this study are shown in Fig.~\ref{fig:fringefields}. As the ratio of the buffer to the drift field is increased, the reach of the leakage field into the drift region increases. In the case of the 10~mm hexagon size, depths of 2~mm or larger are seen for electric field ratios of 2 and above. Both the 2.5~mm and 5~mm hexagons have a leakage field depth of 2~mm or less for all configurations. Several wire widths of 0.13~mm, 0.5~mm, and 1~mm were also studied, and the differences were negligible, leading to a choice of 0.13~mm wire width to maximize optical transparency. The expected operating conditions in NEXT-100 include a buffer electric field of 2~kV/cm and a drift field of 300~V/cm ($E_{\mathrm{buffer}}$/$E_{\mathrm{drift}}$ = 7) giving a leakage field depth of 0.7~mm, 1.3~mm, and 3.5~mm for the 2.5~mm, 5~mm, and 10~mm hexagon sizes respectively. Given the optical transparency of the 5~mm ($\sim$95\%) compared with the 2.5~mm ($\sim$90\%), 5 mm hexagons were chosen for the NEXT-100 cathode.

\subsection{Mesh alignment}

\begin{figure}[t]
\centering
\includegraphics[width=\textwidth]{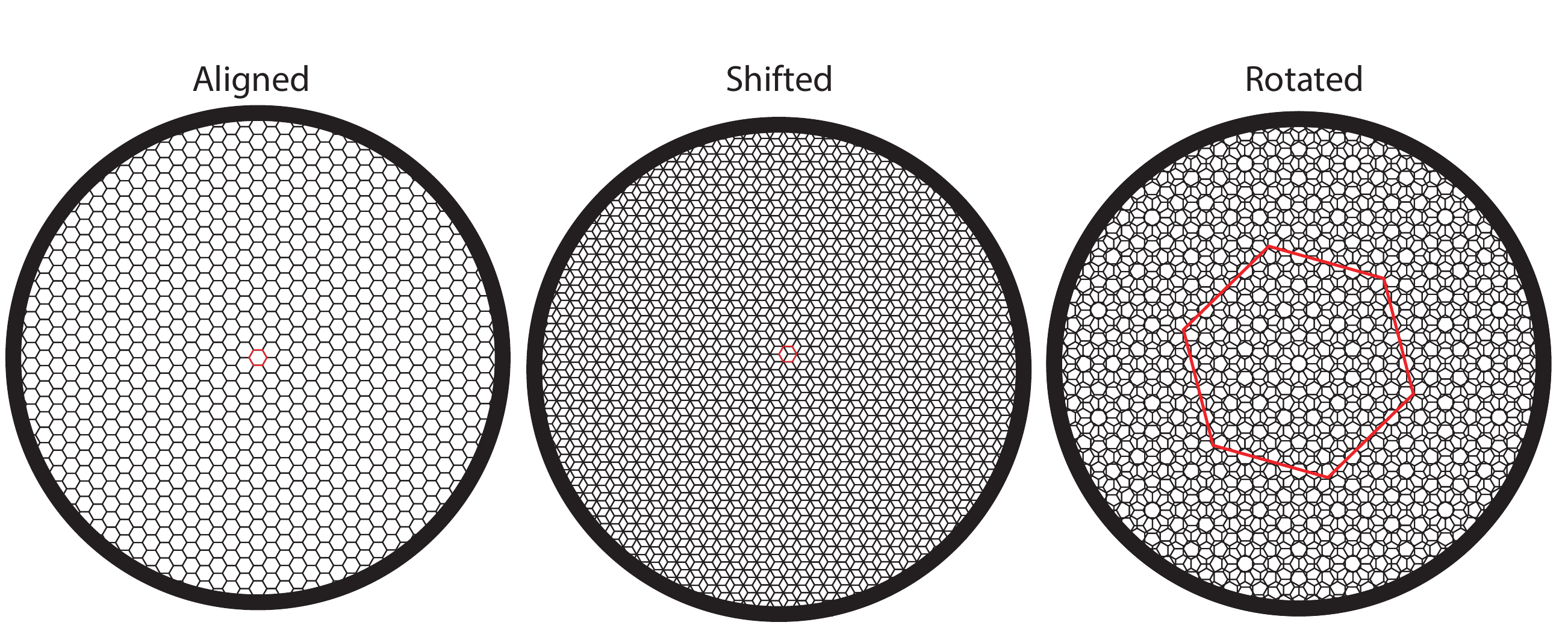}
\caption{\label{fig:mesh_alignment} The mesh alignments studied in this simulation between the gate and anode EL frames as looked at in the $x,y$ plane. Each alignment forms a distinct pattern which alters how the electric field crosses the anode mesh. We consider the maximum 30 degree rotation for the rotated geometry. The red hexagons show example hexagonal unit cells (repeatable structures in the mesh pattern).}
\end{figure}

The relative alignment of the EL gate and anode meshes determines how the electric field lines flow from the gate mesh to the anode, influencing the EL amplification properties of the system. To study this effect we use the same {\tt COMSOL} model for the EL region as shown in Sec.~\ref{sec:comsol} and consider a 60~mm mesh diameter working with the inner 20~mm radius where there is a uniform radial field region (see Fig.~\ref{fig:E_field_radius}). The different mesh alignments considered are shown in Fig.~\ref{fig:mesh_alignment} where we orient the anode mesh such that it gives the desired mesh pattern. 

The electric field lines towards the anode will vary depending on the mesh alignment. A detailed description of the field lines is given in Appendix~\ref{appdix:electric_field}. From these simulations, we find that the transition from the drift region to EL (and from EL to anode) is gradual rather than abrupt and extends several millimeters before and after the mesh surface. In addition, the local electric field around a wire surface is more than a factor of two higher than compared with the field in the EL gap. The different mesh alignments change the location of the hexagon lands at the anode where the field lines can either terminate in these high-field regions at the wire surface or continue to the grounded anode plate.

\begin{figure}[t]
\centering
\includegraphics[width=0.48\textwidth]{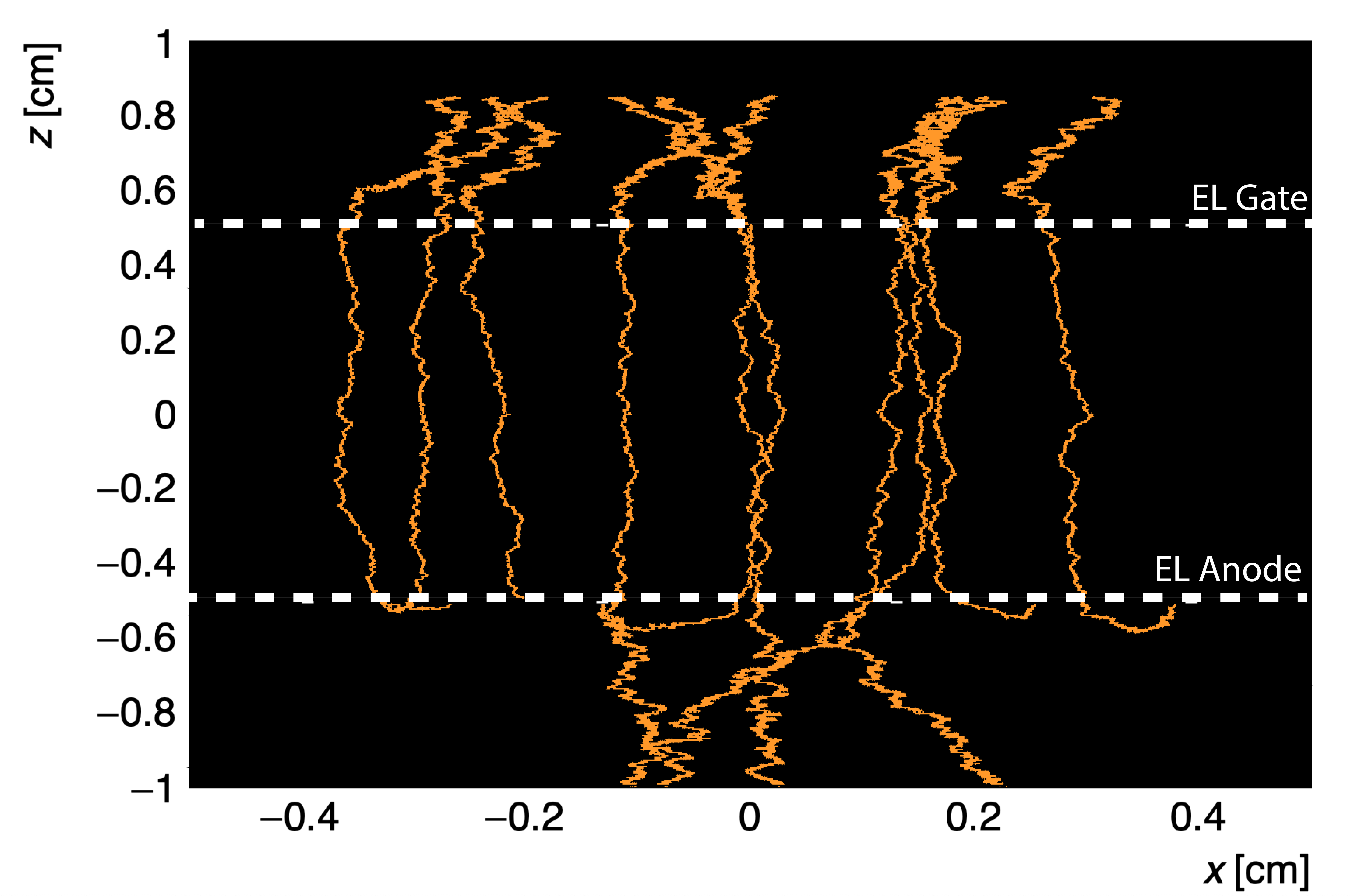}
\includegraphics[width=0.48\textwidth]{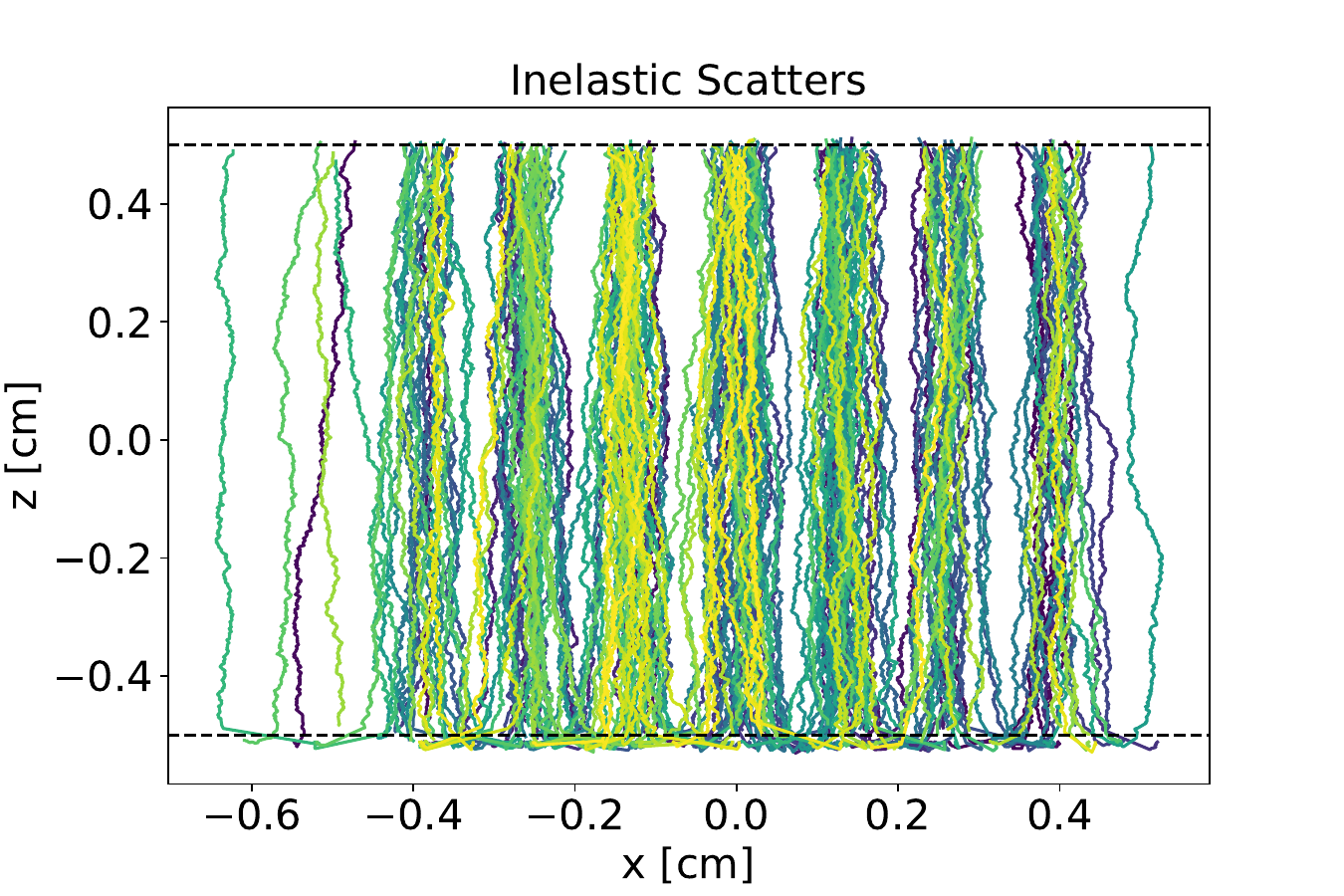}
\caption{\label{fig:garfield} Left: ten electron trajectories simulated in the aligned geometry starting from $z$ = 0.85~cm. Some electrons were collected on the low-field side of the anode mesh, while other electrons were collected on the grounding plate. Right: the electron tracks interpolated from only inelastic scatters (250 electrons are shown). The black dotted lines mark the EL gate ($z$ = 0.5~cm) and EL anode ($z$ = -0.5~cm). Some of the tracks continue to produce light 0.3 mm beyond the anode surface, however, the majority of inelastic scatters that lead to a photon are contained within the EL region. The color scale differentiates different electron trajectories.}
\end{figure}

To study the effect of the electric field non-uniformity on the scintillation photon yield per ionization electron, each of these field maps is imported into a {\tt Garfield++} simulation. Electrons were simulated starting 3.5 mm above the gate mesh and within a radius of 16~mm. {\tt Garfield++} uses a microphysical simulation that steps the electron forward accounting for gas properties, collision cross sections, and diffusion. Gas properties and diffusion constants are calculated via the {\tt MAGBOLTZ} program \cite{magboltz} integrated into the software. The output of the simulation is a series of inelastic scatter locations ($x,y,z,t$) for each electron trajectory.  Each inelastic scatter results in one scintillation photon due to the production of excimers~\cite{SUZUKI1979197}, assuming no impurities in the gas. In the simulations, the field is not high enough to induce any ionization. All studies assume a gas pressure of 13.5 bar and a temperature of 20\textdegree{}C, the expected operating conditions for NEXT-100. Ten example electron trajectories in the $x,z$ plane from the {\tt Garfield++} simulation are shown in Fig.~\ref{fig:garfield}. Electrons follow the field lines towards the center of each hexagon in the gate mesh and get collected on the anode mesh from either the high field region or the grounded side. Some electrons are found to have collected on the grounded plate instead. The locations of all inelastic scatters (leading to photon emission) are also shown in Fig.~\ref{fig:garfield}. Electrons only start to produce inelastic scatters after they cross the gate mesh and stop when they reach a low-field region. The electrons are focused towards the centers of the mesh hexagons, even with random position sampling of their location when starting 3.5 mm above the mesh. Light emission is largely contained within the high-field region. The spreading of the trajectories is due to the diffusion of the electrons.

\begin{figure}[t]
\centering
\includegraphics[width=0.48\textwidth]{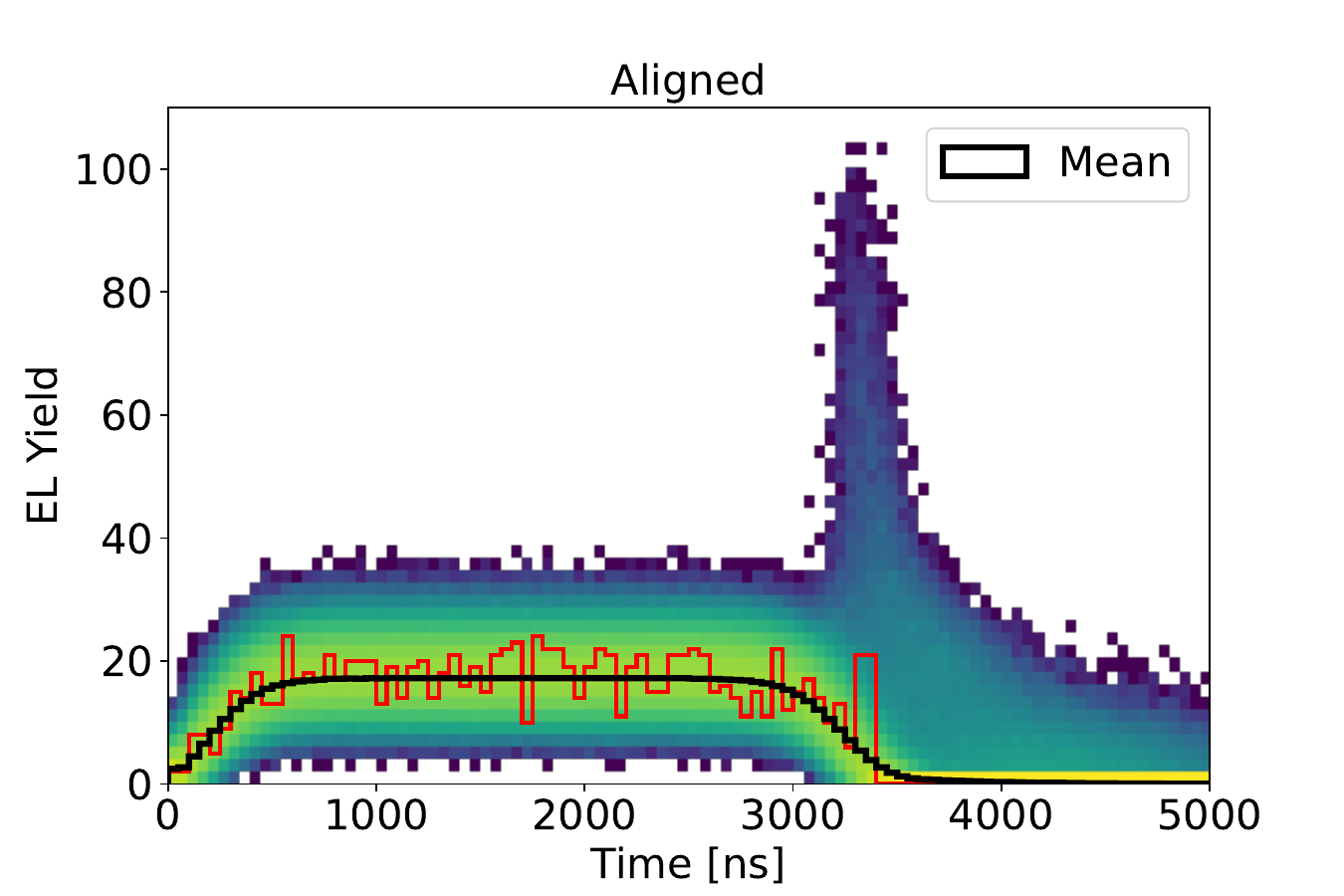}
\includegraphics[width=0.48\textwidth]{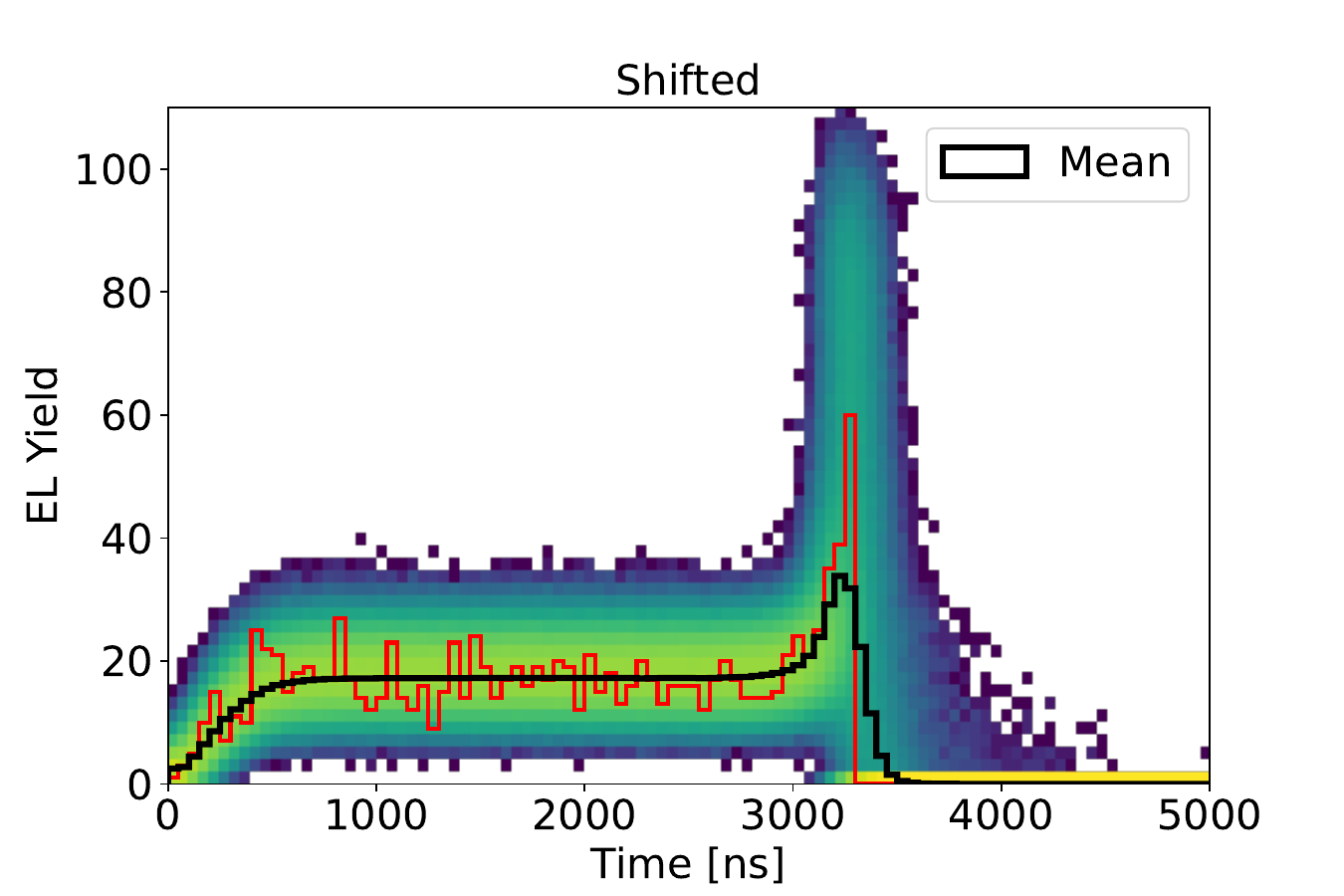}
\includegraphics[width=0.48\textwidth]{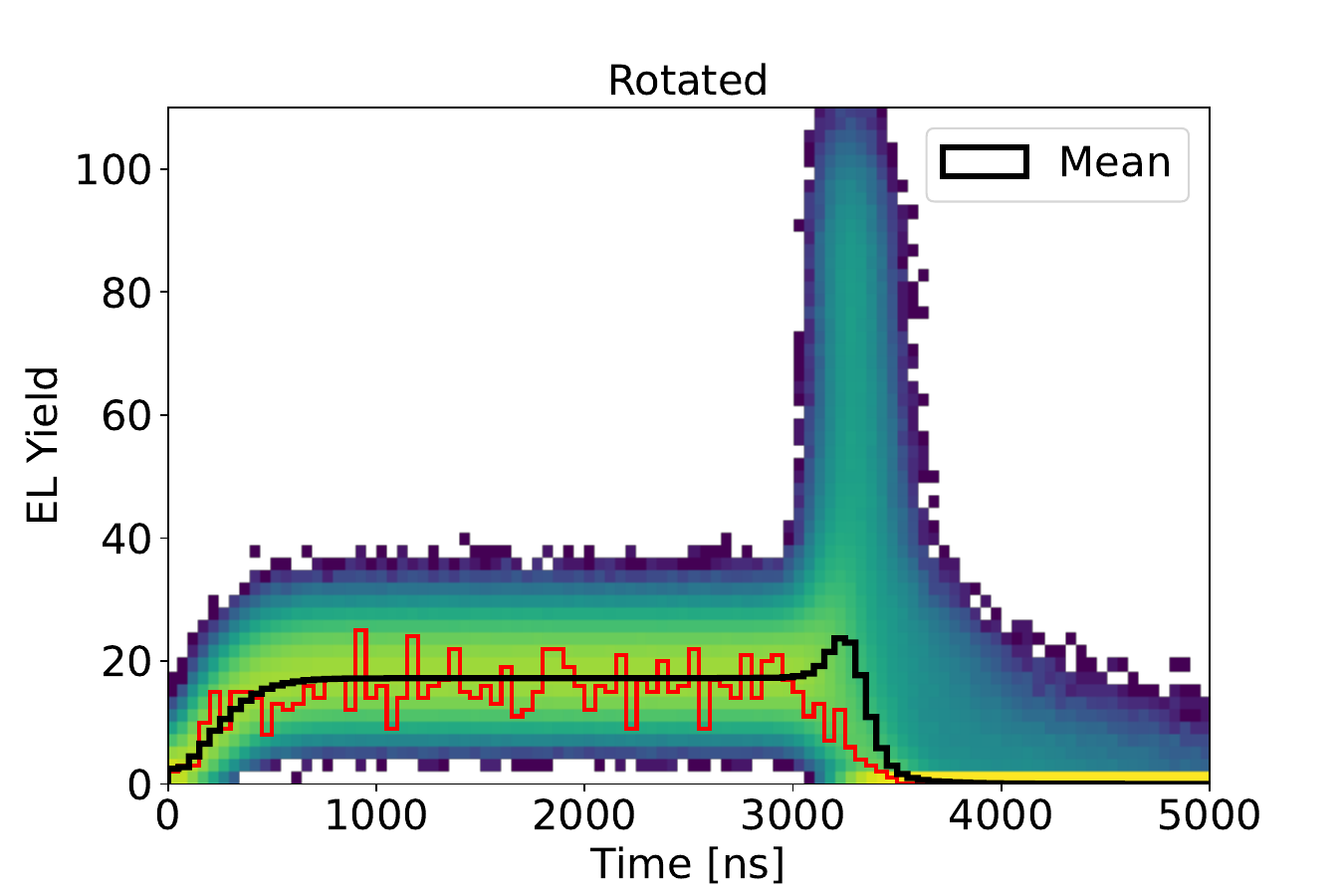}
\includegraphics[width=0.48\textwidth]{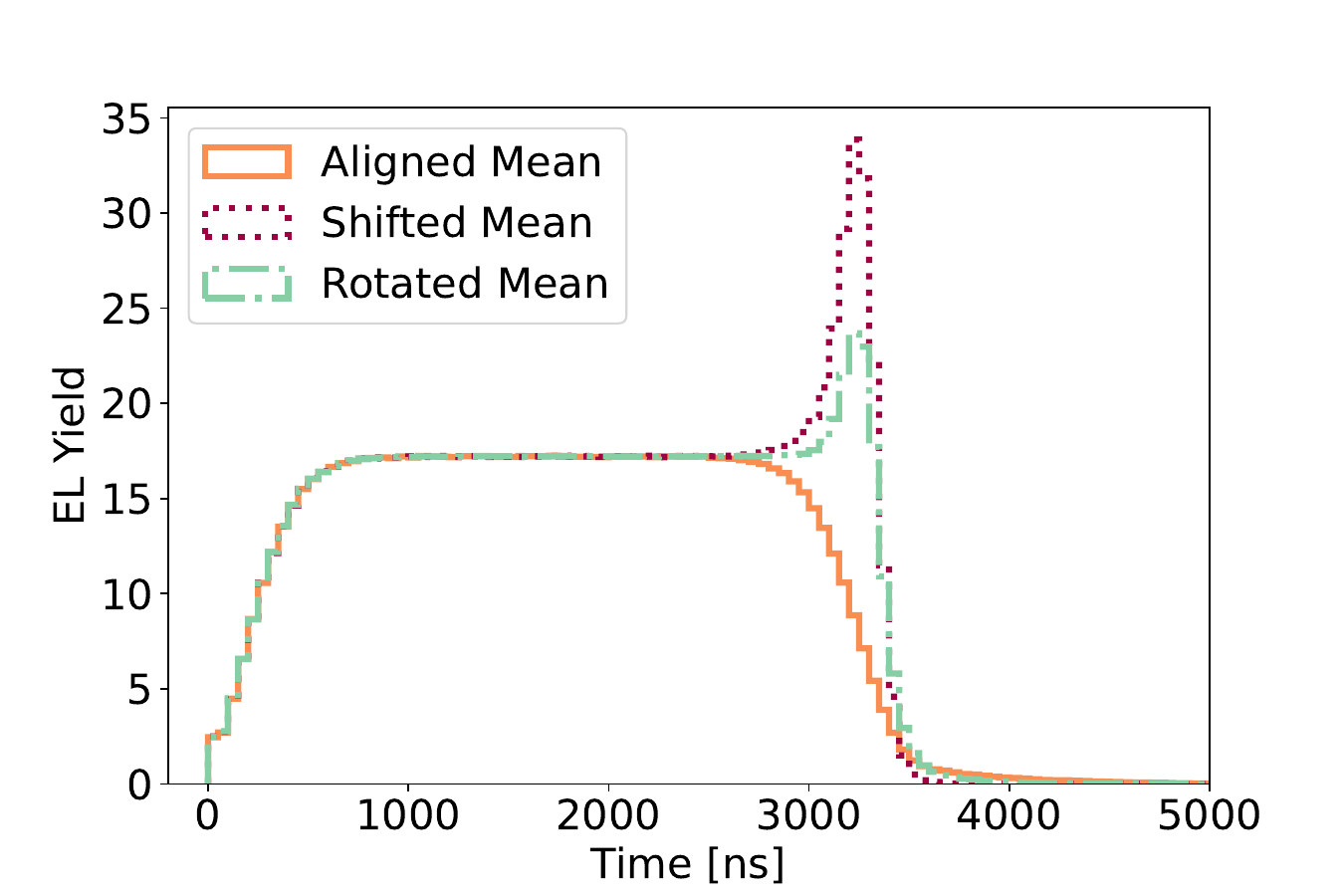}
\caption{\label{fig:time_profiles} Top left, top right, lower left: a 2D histogram of the timing profiles for all electron trajectories in each mesh configuration. The color scale (logarithmic) represents the density of the trajectory in each bin. The red trajectory shows a single electron trajectory and the black line shows the average timing profile. Lower right: The average time profiles overlaid.  }
\end{figure}

The electroluminescent light timing profiles for the simulated electrons are shown in Fig.~\ref{fig:time_profiles}. The mean for each profile is taken from 40,000 simulated electrons. Fluctuations in the timing profile are driven by different path lengths each individual electron takes as it crosses the high field region. The timing profiles largely follow the central electric field line shapes shown in Fig.~\ref{fig:z_field}. In the aligned case, most electrons cross the anode mesh and either continue to the grounding plane or loop back and collect on the low-field side of the anode mesh. These loop-back events manifest as a long tail in the timing profile. In the shifted case, the majority of electrons collect directly on the anode wire surface in the high-field region giving rise to a pulse of light at the end of the path. The rotated timing profile has a mix of aligned and shifted geometries.

\begin{figure}[t]
\centering
\includegraphics[width=0.75\textwidth]{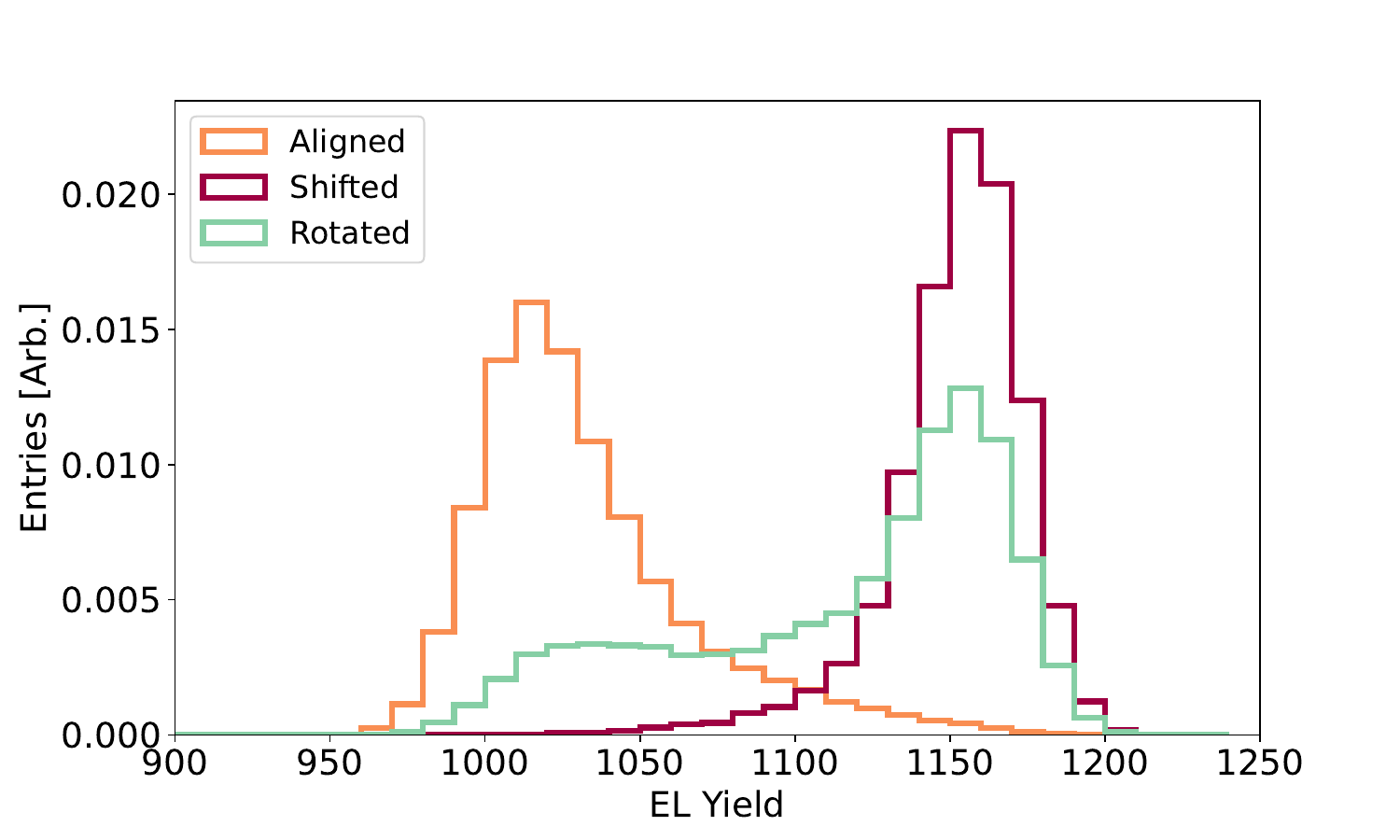}
\caption{\label{fig:EL_yields} The average light yield (area normalized) in a unit cell for each of the mesh configurations at 13.5 bar pressure, 1~cm gap distance, and 20 kV applied to the EL gate.}
\end{figure}

There is an event-by-event variation due to the random path of each electron, variation of the electric field, and stochastic nature of electron-xenon collisions. Figure~\ref{fig:EL_yields} shows a histogram of the total EL yield for each mesh configuration. There is a difference in the EL yield and the spread depending on the configuration. Overall the shifted geometry has the best performance for gain and resolution and the aligned geometry has a lower gain on average by $\sim$10\%. The rotated geometry exhibits behavior that is  intermediate between aligned and shifted geometries. Other rotations were preliminarily explored and were found to similarly represent intermediate cases between aligned and shifted, depending on the angle. 

\subsection{Mesh alignment impact on energy resolution}

To study the impact of the gain variations from mesh alignment on the energy resolution, we assess the impact of these models on events of interest to the NEXT physics program. For this study, a map of EL yields is generated using the 40,000 electrons simulated via {\tt Garfield++} exploiting the hexagonal symmetry of the geometries. For each map, all electrons are translated to a unit cell based on their start $x,y$ position. Example unit cells are highlighted in red in Fig.~\ref{fig:mesh_alignment} where the unit cell for the rotated mesh is much larger than the aligned or shifted geometries. This unit cell is divided into 50 bins using a cubic coordinate system. Each bin contains an array of EL yields to be sampled from. With these maps, we can account for variations in the photon yield from each electron depending on where it enters the gate mesh.

Low-energy (40 keV, 10$^6$ events) and high-energy electrons (2.5 MeV, 10$^5$ events) are then generated in a gaseous xenon volume with dimensions of 0.5 $\times$ 0.4 $\times$ 0.4 m at 13.5 bar (expected operating pressure of NEXT-100) using a {\tt Geant4} \cite{ALLISON2016186,1610988,AGOSTINELLI2003250} simulation and require all energy deposited in the gaseous region. The low-energy events are similar to the calibration samples used in NEXT from $^{83\textnormal{m}}$Kr decays that are effectively point-like, while the high-energy event electrons model the $0\nu\beta\beta$ decay energy of $^{136}$Xe which have longer and more complex topologies. The volume size was chosen to allow for fully contained events and show the effects of diffusion and the angular dependence on tracks to be encapsulated.

\begin{figure}[t]
\centering
\includegraphics[width=0.48\textwidth]{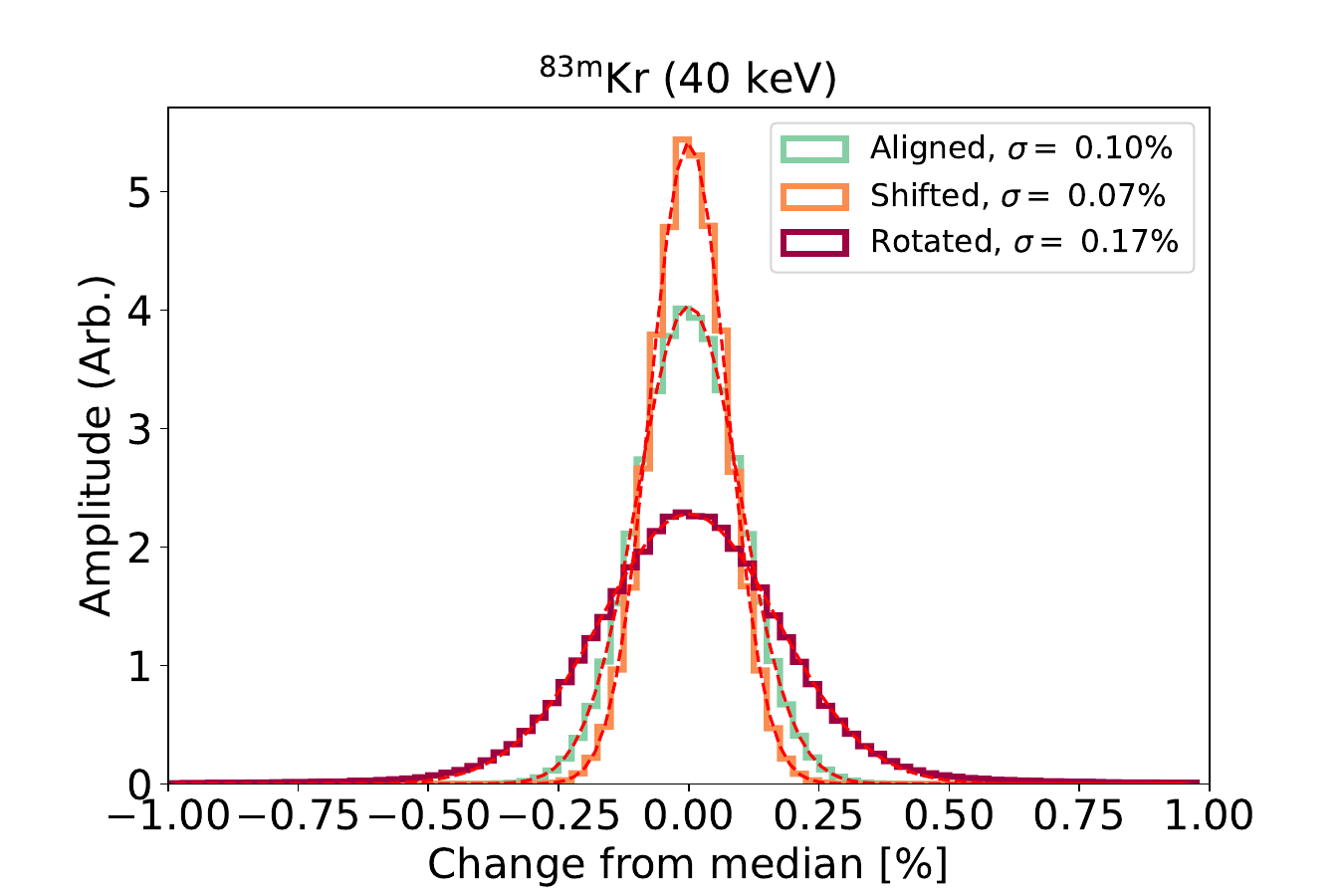}
\includegraphics[width=0.48\textwidth]{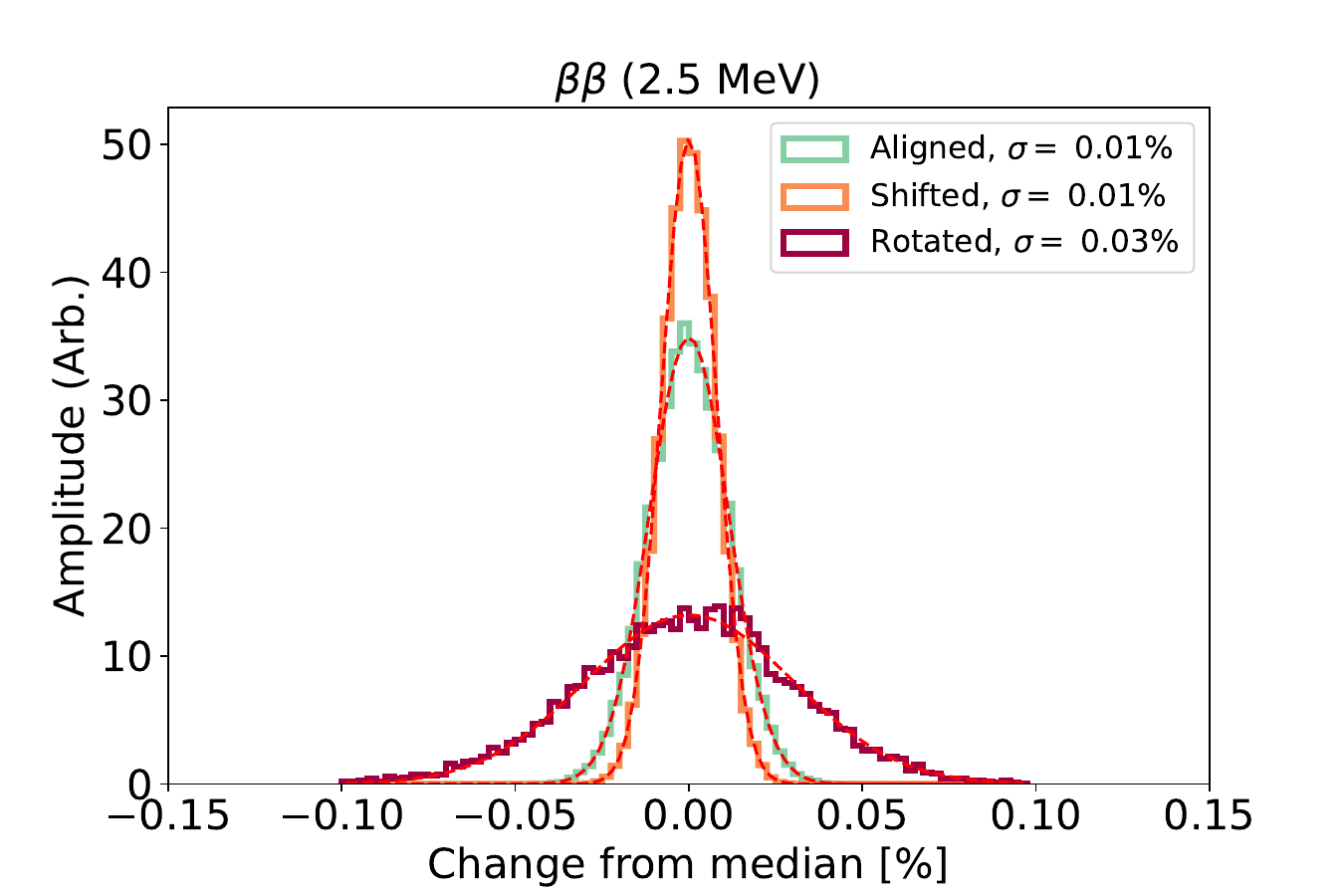}
\caption{\label{fig:Res} The widths in the total light yields produced for each mesh alignment are given in terms of a percent change from the median yield for (left) low and (right) high energy electrons. The dashed red lines are Gaussian fits with the fit sigma given in the legend. Note the different $x$ axis scale for the high-energy electrons.}
\end{figure}

For each event, we take each ionization hit, which contains many electrons at a given $x,y$, and $z$ position inside the volume. The positions of each electron in a hit are smeared using a two-dimensional Gaussian distribution to account for transverse diffusion. This smearing models the diffusion of the ionization electrons towards a readout plane located at the end of the TPC. For each ionization electron, each smeared $x,y$ coordinate is translated to a position in the unit cell on the EL yield map, and a light yield is randomly sampled based on the bin it is located. The total yield is summed for all ionization electrons and is binned in a histogram. The histogram is then fit with a Gaussian distribution to determine the spread in light yield for a given event. The results are summarized in Fig.~\ref{fig:Res} and show that each mesh configuration at low or high energy has a sub-percent effect on the total light yields produced. A lower operating voltage of 15 kV was also checked for each alignment, we found that the impact on energy resolution was slightly larger but within 0.5\% and 0.1\% for low-energy and high-energy events respectively. 

The larger impact on lower energy events is due to the size of the events being more similar to the size of the hexagons. The rotated mesh has a wider variation due to the larger variation in path lengths possible due to the geometry. We conclude that any mesh alignment is suitable for use in NEXT-100. Due to existing design constraints and simplicity of assembly, the meshes in NEXT-100 were oriented with a rotation of approximately 15 degrees. As part of the NEXT-100 commissioning process, we plan to verify the waveform shapes using data from $^{83\textnormal{m}}$Kr decays and further study the edge effects of the electric field. 


\section{Installation of EL and cathode regions in NEXT-100}\label{Sec:installation}
\begin{figure}[t]
\centering
\includegraphics[width=\textwidth]{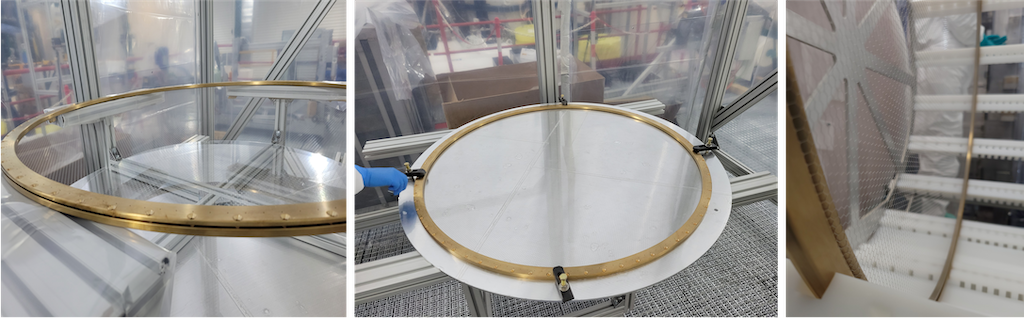}
\caption{\label{fig:cathode_assembly} Images of various stages of the cathode being assembled. Left: image of the tensioning ring being placed on top of the mesh and base ring just before tensioning. Middle: the cathode being clamped down while it is tensioned. Right: an image of the cathode being inserted into the field cage. }
\end{figure}

Construction of the EL and cathode regions and installation in the NEXT-100 detector was carried out at the LSC in June 2023. The cathode is one of the first components installed in the field cage. Figure \ref{fig:cathode_assembly} shows three stages of the cathode construction. The left picture shows the tensioning ring being placed on top of the mesh and base ring before any screws have been added. The middle picture shows the cathode rings clamped to a table during tensioning of the mesh. The right picture shows a close-up image of the cathode that was placed between the white HDPE staves. The staves provide the structure of the field cage for the insertion of the copper field rings and Teflon reflector panels and insulate between each field ring.

\begin{figure}[t]
\centering
\includegraphics[width=\textwidth]{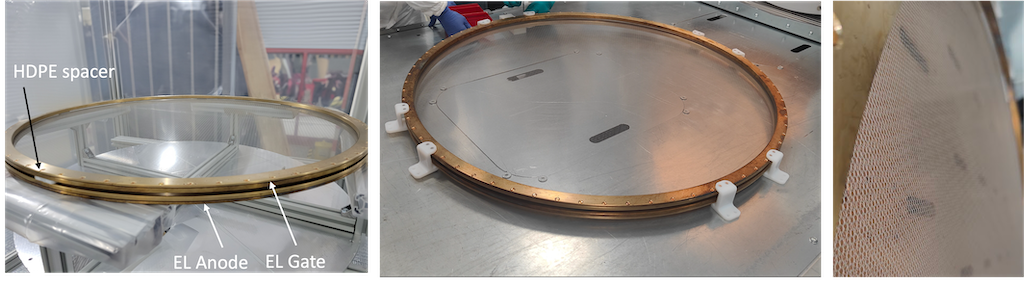}
\caption{\label{fig:NEXT100} Left: a picture of the NEXT-100 EL region before joining them together with the brackets. Middle: the EL region with the HDPE brackets attached just before placing on the copper end cap. Right: a close-up picture of the meshes in the EL region. }
\end{figure}

Figure \ref{fig:NEXT100} shows pictures of various construction stages of the NEXT-100 EL region. An image of the EL frames prior to adding the brackets is shown. The white spacers between the rings separate them such that the mesh surfaces do not touch before adding the brackets. The middle picture shows the EL rings with the HDPE brackets just before they were placed onto the copper tracking plane. The final picture shows a close-up image of the EL gate. The final EL meshes following tensioning include one mesh without breaks and one mesh with three breaks. The mesh with the breaks was placed at the anode which reduces the formation of the sparks at these points. 

The EL region has been tested in 4 bar of nitrogen (which is similar to $\sim$15 bar xenon estimated from {\tt PyBoltz}~\cite{ALATOUM2020107357,Norman_2022}) and reached a breakdown near the center of the mesh at an electric field strength of 17.4~kV/cm which meets the minimum operating specifications for NEXT-100.

The construction of the remaining parts of NEXT-100 including tracking and energy planes is now proceeding. The full detector will be exercised beginning in Winter 2023, with a physics program to follow a short commissioning phase that is expected to run to Summer 2024.

\section{Conclusion}

In this work, we have described the design choices made for the NEXT-100 electroluminescent and cathode region which have been mechanically installed for first tests of operation in the detector located at the LSC in Spain. We studied the electrical robustness of the hexagonal meshes to sparks and investigated the electric breakdown potential and positioning in air. We found that the breakdown tends to happen near the center of the mesh where the deflection is larger and occasionally breakdown occurs near a defective position (such as a break) on the mesh surface. An analytical model was developed to predict the electrostatic deflection with and without a support post placed at the center of the meshes to reduce the deflection. Measurements of the electrostatic deflection were performed by bringing a segment of the mesh into focus using a microscope objective and a webcam. The system is positioned on a micrometer stage to achieve 8~\si{\micro\meter} precision at a given point. Measurements were taken of the deflection as a function of mesh tension and we found that the deflection is reduced as more tension is applied to the mesh. Finally, measurements of the NEXT-100 EL region were carried out before its installation at the LSC with extracted tension values to be 990$\pm$45 N and 835$\pm$40 N for each mesh. These measurements provide the expected deflection for the operation of the NEXT-100 EL region at a given electric field. Simulations of the electric field of the EL and cathode are performed including studies of the effect of different alignments of the meshes in the EL region including aligning, shifting, or rotating them by 30 degrees. The simulations include a detailed CAD model of the EL region with the {\tt COMSOL Multiphysics} program to calculate the electric field. This field is further used in a {\tt Garfield++} simulation to calculate the EL gain for individual electrons crossing the gap in each alignment. We found that the mesh alignment does not significantly affect the energy resolution of low-energy (40 keV) and high-energy (2.5 MeV) events. Finally, we reported on the assembly of the NEXT-100 EL and cathode at the LSC, and installation into the detector. With the TPC now fully installed, NEXT-100 will proceed to first full detector commissioning runs.

\section*{Acknowledgments}
The NEXT Collaboration acknowledges support from the following agencies and institutions: the European Research Council (ERC) under Grant Agreement No. 951281-BOLD; the European Union’s Framework Programme for Research and Innovation Horizon 2020 (2014–2020) under Grant Agreement No. 957202-HIDDEN; the MCIN/AEI of Spain and ERDF A way of making Europe under grants PID2021-125475NB and the Severo Ochoa Program grant CEX2018-000867-S; the Generalitat Valenciana of Spain under grants PROMETEO/2021/087 and CIDEGENT/2019/049; the Department of Education of the Basque Government of Spain under the predoctoral training program non-doctoral research personnel; the Spanish la Caixa Foundation (ID 100010434) under fellowship code LCF/BQ/PI22/11910019; the Portuguese FCT under project UID/FIS/04559/2020 to fund the activities of LIBPhys-UC; the Israel Science Foundation (ISF) under grant 1223/21; the Pazy Foundation (Israel) under grants 310/22, 315/19 and 465; the US Department of Energy under contracts number DE-AC02-06CH11357 (Argonne National Laboratory), DE-AC02-07CH11359 (Fermi National Accelerator Laboratory), DE-FG02-13ER42020 (Texas A\&M), DE-SC0019054 (Texas Arlington) and DE-SC0019223 (Texas Arlington); the US National Science Foundation under award number NSF CHE 2004111; the Robert A Welch Foundation under award number Y-2031-20200401. Finally, we are grateful to the Laboratorio Subterr\'aneo de Canfranc for hosting and supporting the NEXT experiment.

\appendix

\section{Analytical calculation of the electrostatic deflection}\label{appdix:deflection}
\begin{figure}[t]
\centering
\includegraphics[width=0.3\textwidth]{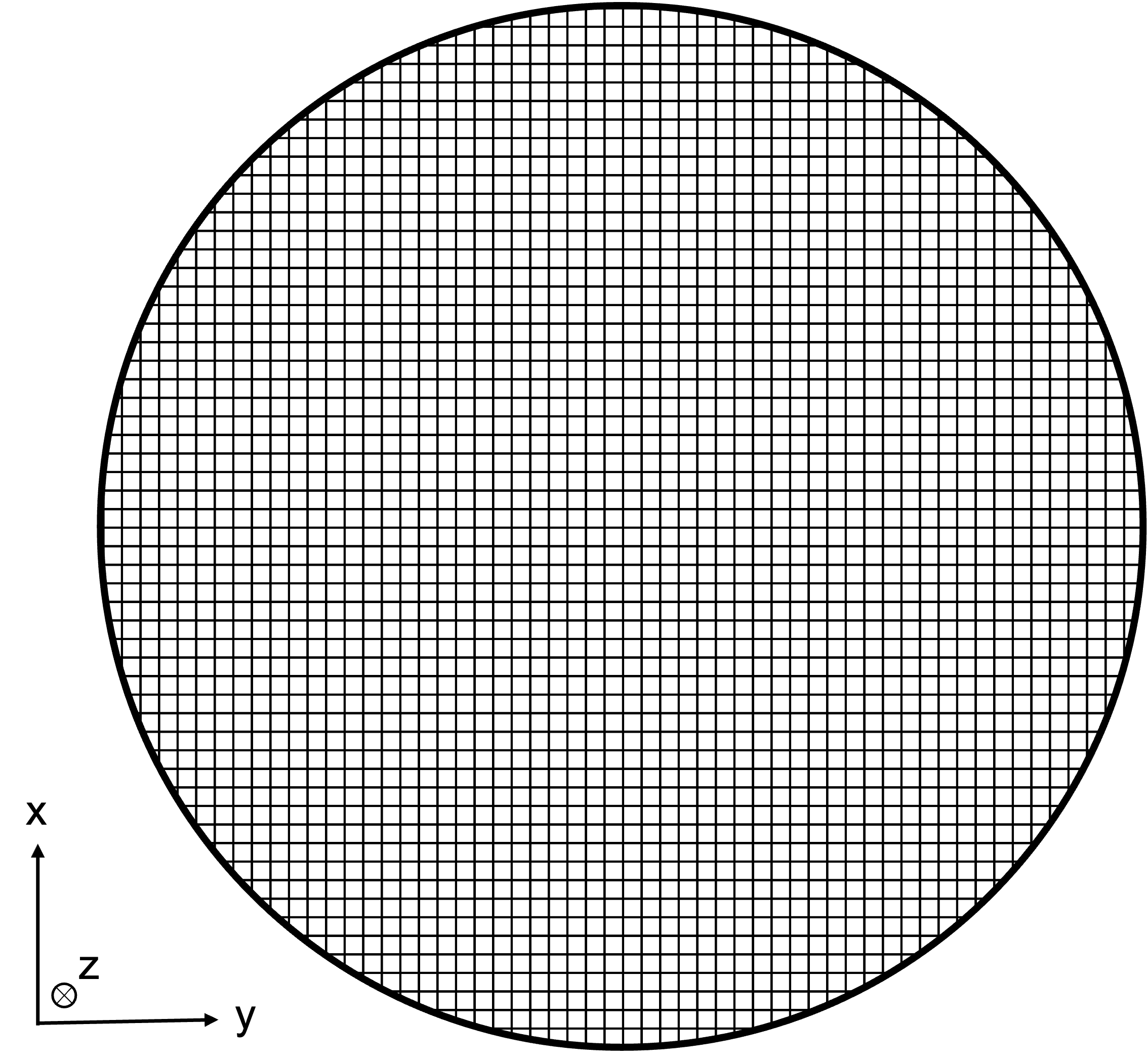}
\caption{\label{fig:mesh_co_sys} The coordinate system used for calculating the expected electrostatic deflection of an EL mesh. The second mesh will be displaced from the mesh shown in the $z$-direction. The origin is taken to be the center of the mesh which is circularly symmetric. }
\end{figure}

This appendix gives the full calculation for the electrostatic deflection of a circular mesh when charged under a voltage. The calculation considers one mesh to be charged to a voltage, $V$ (known as the gate mesh), and the other mesh to be at ground (the anode mesh) - akin to the geometry of a parallel plate capacitor. For simplicity, the mesh is modeled as a series of parallel wires spanning the $x$ and $y$ directions. The coordinate system used is shown in Fig.~\ref{fig:mesh_co_sys}. 

The stored energy in each segment of the mesh has contributions from the elastic energy due to the tensioning of the mesh and from the electrostatic energy from the capacitance between the mesh surface. The equilibrium shape is determined by minimizing the sum of these contributions using the Euler-Lagrange equation. The calculations are detailed in what follows.  
\subsection*{Elastic energy}
For an EL mesh that is pre-tensioned to a tension, $T$, a wire segment in this mesh of initial length $\delta l_0$ stores elastic potential energy (in zero-field), $U^0_{\textrm{elastic}}$, given by:

\begin{equation}
    U^0_{\textrm{elastic}} = \frac{1}{2}\frac{A\delta l_0}{Y}\sigma^2,
\end{equation}
where $A$ is the cross section of a wire segment in the mesh, $Y$ is the Young's modulus and $\sigma$ is the stress. We can rewrite this in terms of the tension, $T= \sigma A$:
\begin{equation}
    U^0_{\textrm{elastic}} = \frac{1}{2}\frac{\delta l_0}{YA}T^2.
\end{equation}
The length of the pre-tensioned segment at zero field is given by:
\begin{equation}\label{deltapt}
    \delta l_{\textrm{pt}} = \delta l_0 \Big( 1 + \frac{T}{AY}\Big)
\end{equation}

With a voltage applied, the segment will stretch further. The deformed length, $\delta l_{\textrm{def}}$ can be related to the un-deformed length by:
\begin{equation}
    \delta l_{\textrm{def}} =  \delta l_{\textrm{pt}} \Bigg [ \sqrt{1 + \Big (\frac{dz}{dx} \Big  )^2} \Bigg ],
\end{equation}
where $dz/dx$ is the deformation gradient. Assuming the gate-to-anode mesh separation in $z$ is much smaller than the radius of the mesh, we can assume $dz/dx\ll1$, and expand to first order:
\begin{equation}\label{deltadef}
    \delta l_{\textrm{def}} =  \delta l_{\textrm{pt}} \Bigg [ 1 + \frac{1}{2}\Big (\frac{dz}{dx} \Big  )^2 \Bigg ].
\end{equation}

The stored energy in a single wire is given by:
\begin{equation}
    U_{\textrm{elastic}} = \frac{AY\delta l_0}{2} s^2 = \frac{AY\delta l_0}{2} \Bigg( \frac{\delta l_{\textrm{def}} - \delta l_0}{\delta l_0} \Bigg )^2,
\end{equation}
where $s$ is the strain. Using Eqns. \ref{deltapt} and \ref{deltadef} and rearranging, we find:
\begin{align}
    U_{\textrm{elastic}} = \frac{AY\delta l_0}{2} \Bigg(  \frac{T}{AY} + \Big( 1 + \frac{T}{AY} \Big) \Big ( \frac{dz}{dx} \Big )^2 \Bigg )^2
    \\
    = \frac{T\delta l_0}{2} \Bigg [  \Big (\frac{T}{AY} \Big ) + \Big( 1 + \frac{T}{AY} \Big) \Big ( \frac{dz}{dx} \Big )^2 \Bigg ] + \mathcal{O}\Big (\frac{dz}{dx} \Big )^4
\end{align}
The total stored energy is obtained by integrating over all wires. A wire along the $x$-direction at position $y$ has a tensioned length that stretches from $-x_0 = -\sqrt{R^2 - y^2}$ to $x_0 = \sqrt{R^2 - y^2}$, where $R$ is the radius of the mesh. Integrating along the pre-tensioned wire is equivalent to identifying the integration measures $dx=\delta l_{\textrm{pt}}$,
\begin{equation}
    U_{\textrm{elastic}}^{1~\textrm{wire}} = \int^{\sqrt{R^2 - y^2}}_{-\sqrt{R^2 - y^2}}dx\frac{T}{2} \Bigg [ \Big( \frac{T}{AY} \Big ) \Big( 1 + \frac{T}{AY}\Big)^{-1} + \Big ( \frac{dz}{dx} \Big )^2  \Bigg ].
\end{equation}

If there are $N_x$ wires per unit length in $y$, we sum to find the contributions from all $x$ wires:
\begin{equation}
    U_{\textrm{elastic}}^{x~\textrm{wires}} = \sum^{j\sim RN_x}_{j\sim -RN_x} \int^{R\sqrt{1 - j^2}}_{-R\sqrt{1 - j^2}}dx\frac{T}{2} \Bigg [ \Big( \frac{T}{AY} \Big ) \Big( 1 + \frac{T}{AY}\Big)^{-1} + \Big ( \frac{dz}{dx} \Big )^2  \Bigg ].
\end{equation}
If the wires are sufficiently densely packed, we can replace the sum with an integral:

\begin{equation}
    \sum^{j\sim RN_x}_{j\sim -RN_x} \int^{R\sqrt{1 - j^2}}_{-R\sqrt{1 - j^2}}dx \to \int^{RN_x}_{-RN_x} dj \int^{R\sqrt{1 - j^2}}_{-R\sqrt{1 - j^2}}dx.
\end{equation}

Writing $j/N_x = y$, 
\begin{equation}
    N_x \int^{R}_{-R}dy  \int^{\sqrt{R^2 - y^2}}_{-\sqrt{R^2 - y^2}}dx = N_x\int_{\circ R}dxdy, 
\end{equation}
thus, the total elastic contribution from the $x$ wires is:
\begin{align}
    U_{\textrm{elastic}}^{x~\textrm{wires}} = N_x\int_{\circ R}dxdy\frac{T}{2}\Bigg [ \Big( \frac{T}{AY} \Big ) \Big( 1 + \frac{T}{AY}\Big)^{-1} + \Big ( \frac{dz}{dx} \Big )^2  \Bigg ]
    \\
    = U_0^x + N_x\int_{\circ R}dxdy\frac{T}{2}\Big ( \frac{dz}{dx} \Big )^2.
\end{align}
Adding the similar contribution from the $y$ wires, we find the total stored elastic energy is given by:
\begin{equation}
    U_{\textrm{elastic}} = U_0^x + U_0^y + \int_{\circ R}dxdy\frac{T}{2} \Bigg [ N_x\Big ( \frac{dz}{dx} \Big )^2 + N_y\Big ( \frac{dz}{dx} \Big )^2 \Bigg ].
\end{equation}
Using this formula, we can see that if the wire was not pre-tensioned, the only restoring force would come from the $\mathcal{O}(\frac{dz}{dx})^4$ term, which was neglected here. The mesh will always be pre-tensioned for our purposes, so the second-order term suffices. 

\subsection*{Electrostatic energy}
We can assess the total capacitive energy by assuming each surface element contributes as a small parallel plate capacitor and all are connected in parallel. Thus, the total electrostatic energy is given by:
\begin{equation}
    U_{\textrm{elec}} = \frac{1}{2}CV^2 = \sum_{i}\frac{1}{2}C_iV^2 = \sum_{i}\frac{1}{2}\epsilon \frac{dxdy}{g+z}V^2,
\end{equation}
where $C$ is the capacitance, $g$ is the gap distance between the meshes and $V$ is the voltage. Taking the infinitesimal limit:

\begin{equation}
    U_{\textrm{elec}} \to  \int_{\circ R} dxdy \frac{1}{2}\epsilon \frac{V^2}{g+z}
\end{equation}

We will generally be able to assume small displacements of the mesh, although this is not as robust an approximation as the ones taken previously in this calculation, however, it will provide a considerable simplification:
\begin{equation}
    U_{\textrm{elec}} \sim  \int_{\circ R} dxdy \frac{1}{2}\epsilon \frac{V^2}{g}\Big (1 - \frac{z}{g} \Big)
\end{equation}
\subsection*{Equilibrium shape}\label{Apdx:eqshape}
To find the equilibrium mesh shape $z(x,y)$, we need to minimize the energy functional:
\begin{align}
    U =  U_0^x + U_0^y + \int d^2x_i\varepsilon(z)
    \\
    \varepsilon(z) = \frac{T}{2} \Bigg [ N_x\Big ( \frac{dz}{dx} \Big )^2 + N_y\Big ( \frac{dz}{dy} \Big )^2 \Bigg ] + \frac{1}{2}\epsilon \frac{V^2}{g}\Big (1 - \frac{z}{g} \Big),
\end{align}
which can be solved with the Euler-Lagrange equation:
\begin{equation}
    \sum_{i}\partial_i\Big( \frac{\partial\varepsilon}{\partial[\partial_i z]} \Big) = \frac{d\varepsilon}{dz}.
\end{equation}
Applying the Euler-Lagrange equation, we find,
\begin{equation}
    T\Big( N_x\frac{\partial^2z}{\partial x^2} + N_y\frac{\partial^2z}{\partial y^2} \Big) = -\frac{1}{2}\epsilon\frac{V^2}{g^2}.
\end{equation}
If we restrict to the symmetric case, $N_x = N_y = N$, in which case, this reproduces to a cylindrically symmetric form:
\begin{equation}\label{eq:EqA23}
    (\partial^2_x + \partial^2_y)z = -4\kappa, \quad \kappa = \frac{1}{4}\cdot \frac{\epsilon E^2}{2TN},
\end{equation}
where we have substituted in the electric field, $E = V/g$. The additional factor of $1/4$ is introduced for convenience in the calculations. Solving using cylindrical polar coordinates, we can rewrite Eqn.~\ref{eq:EqA23}:
\begin{equation}\label{eq:polar}
    \frac{1}{\rho}\frac{\partial}{\partial\rho}\Big( \rho\frac{\partial}{\partial\rho}\Big)z = -4\kappa.
\end{equation}
To find the general solution, we write $z$ as a power series in $\rho$:
\begin{equation}
    z = \sum_{n}a_n\rho^n,
\end{equation}
which, when substituted into Eqn.~\ref{eq:polar}, gives:
\begin{equation}\label{eq:polarsol}
    \sum_{n}n^2a_n\rho^{n-2} = -4\kappa.
\end{equation}
Since there are no terms with $\rho$ on the right-hand side of Eqn.~\ref{eq:polarsol}, only the $n=0$ and $n=2$ terms contribute. The boundary condition $z(R)=0$ fixes the relationship between $a_0$ and $a_2$ to yield:
\begin{equation}
    a_0 = \frac{\kappa}{R^2}, \quad a_2 = -\kappa, \quad a_{\textrm{others}}=0.
\end{equation}
This gives the solution:
\begin{equation}
    z = -\kappa(R^2 - \rho^2)
\end{equation}
We can express $\kappa$ in terms of the total mesh tension, $\tau=2TNR$, rather than the tension per wire, $T$, as:
\begin{equation}
    \kappa = \frac{\epsilon R E^2}{4\tau}.
\end{equation}
The extremal electrostatic deflection occurs at $\rho = 0$ and is:
\begin{equation}
\label{eq:y:3}
z^{\textrm{max}} = -\kappa R^2.
\end{equation}
It is useful to note that the electrostatic deflection does not depend on the Young's modulus of the mesh. 
\subsection*{Addition of a support post}
In the scenario of adding an insulative support post between the meshes to reduce the deflection, we can use a similar method and solve the Euler-Lagrange equations. In this case, we have slightly different boundary conditions  for which we want to solve Eqn.~\ref{eq:polar} due to the post situated at the origin. Again using cylindrical polar coordinates and solving the equation with integration exactly instead of using a power series:
\begin{equation}\label{eq:intsol}
    z = -\kappa \rho^2 + A\ln{\rho}+B,
\end{equation}
where $A$ and $B$ are integration constants that we need to determine. For any solution going to $\rho = 0$, we will have to throw the logarithmic term by setting $A = 0$. This case gives us back the polynomial solution we found in Sec.~\ref{Apdx:eqshape}. If we don't try to find the solution all the way to the origin, we can consider a boundary condition of $z = 0$ at $r = R_0$ and $r = R$ and find the displacement anywhere between these values. The value of $R_0$ is the effective diameter of the post which includes the width of the post and the region of mesh which is stiff enough not to collapse around the head of the post.

It is more convenient to write Eqn. \ref{eq:intsol} in the following form:
\begin{equation}
    z = -\kappa \rho^2 + A \ln{\frac{\rho}{C}},
\end{equation}
where $A$ and $C$ both have units of length. Our two boundary conditions read:
\begin{equation}\label{eq:intconst}
    0 = -\kappa R^2 + A \ln{\frac{R}{C}}, \\ 0 = -\kappa R_0^2 + A \ln{\frac{R_0}{C}}.
\end{equation}
Using the right-hand equation to solve for $C$, we find:
\begin{equation}
    C = R_0\exp{-\frac{\kappa R_0^2}{A}}.
\end{equation}
Substituting this equation into the Eqn.~\ref{eq:intconst} (left) and rearranging, we find:
\begin{equation}
    A = \frac{\kappa (R^2 - R_0^2)}{\ln({R/R_0})},
\end{equation}
for which we can then determine $C$:
\begin{equation}
    C = R_0 \Big ( \frac{R}{R_0} \Big )^{-\frac{R_0^2}{R^2 - R_0^2}}.
\end{equation}
We can then write the full expression for the deflection with the post as:
\begin{equation}\label{eq:def_post}
    z = -\kappa\rho^2 + \kappa \Bigg ( (R^2 - R_0^2)\frac{\ln(\rho/R_0)}{\ln(R/R_0)} + R_0^2  \Bigg ).
\end{equation}

To find the maximum deflection, we differentiate Eqn.~\ref{eq:def_post} and set it equal to zero and solve:
\begin{equation}
    \rho_{\textnormal{max}} = \sqrt{ \frac{R^2 - R_0^2}{2\ln(R/R_0)} },
\end{equation}
which gives:
\begin{equation}
    z_{\textnormal{max}} = \kappa(R_0^2 - R^2)\Bigg [ \frac{\ln \Bigg ( \frac{R^2 - R_0^2}{2R^2\ln(R/R_0)} \Bigg ) - 1}{2\ln(R/R_0)} \Bigg ] + \kappa R^2.
\end{equation}

\section{Electrostatic deflection measurements with a support post}\label{appdix:deflection_measurements}

\begin{figure}[t]
\centering
\includegraphics[width=0.48\textwidth]{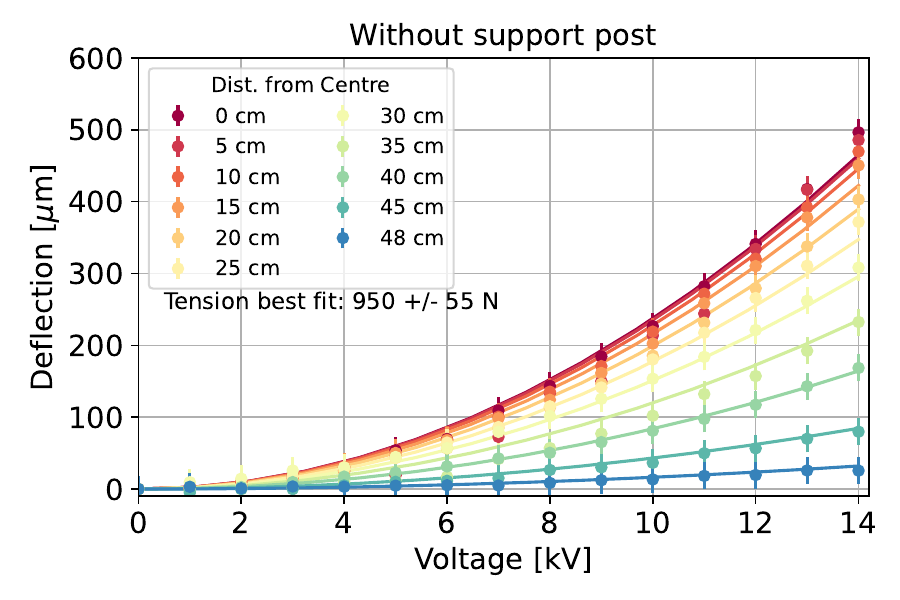}
\includegraphics[width=0.48\textwidth]{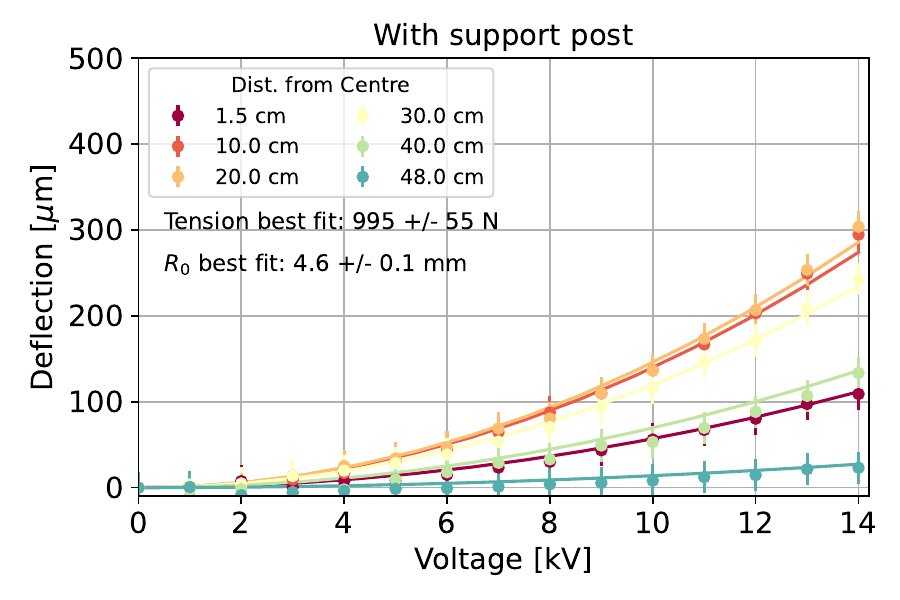}
\includegraphics[width=0.285\textwidth]{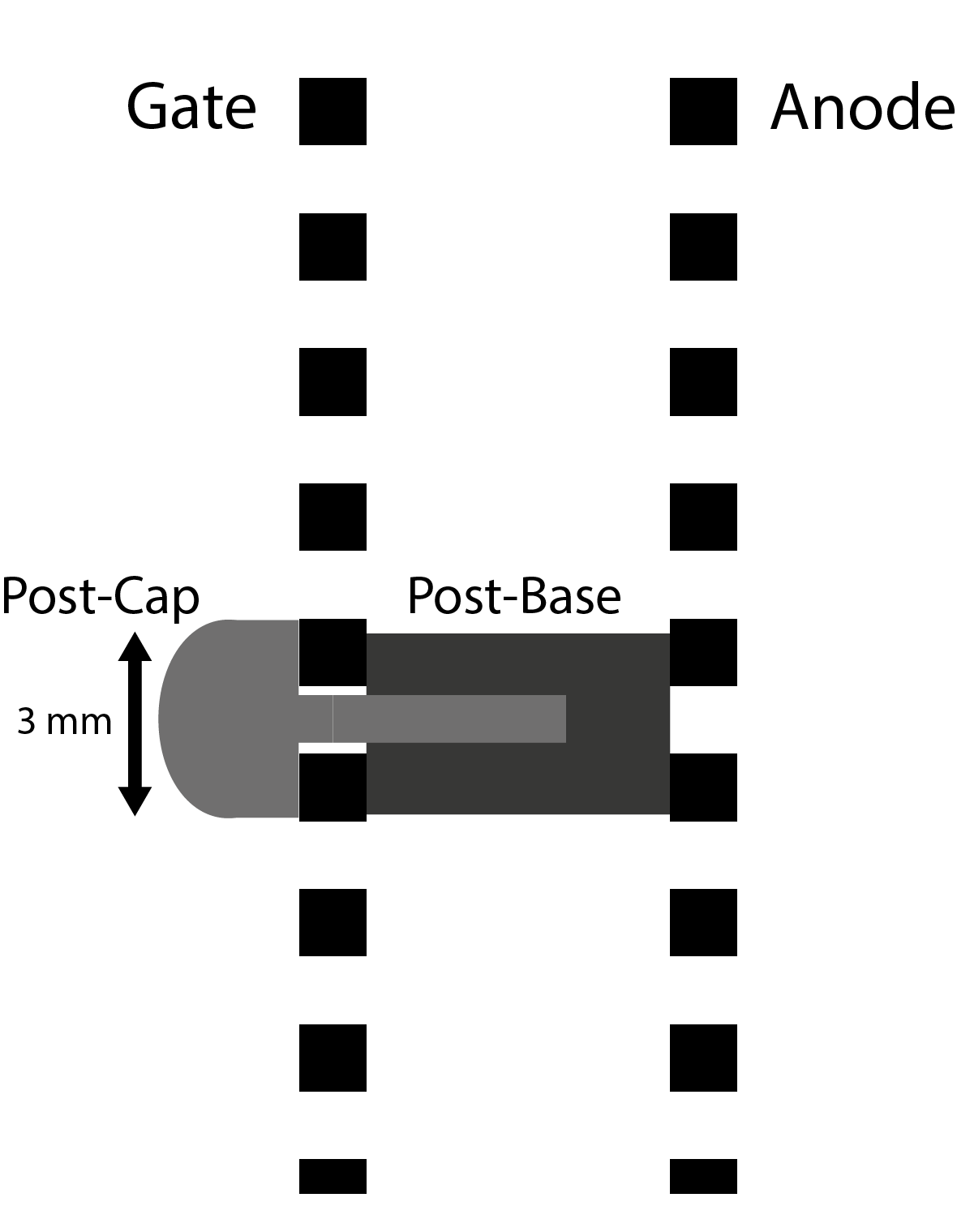}
\includegraphics[width=0.48\textwidth]{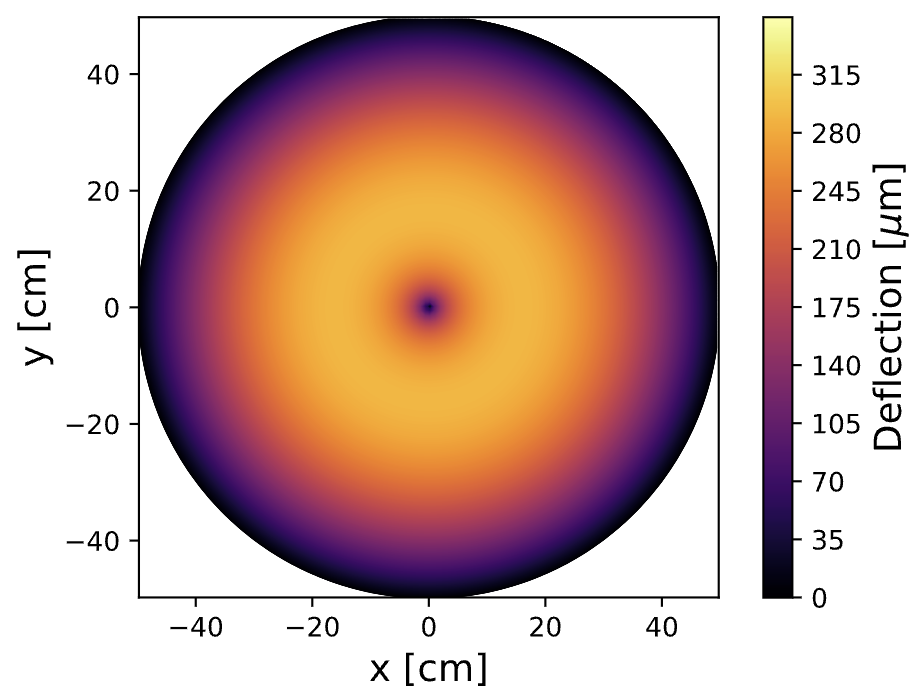}

\caption{\label{fig:def_magic} Top left: the electrostatic deflection without a support post. Top right: the electrostatic deflection with a support post between the meshes. Bottom left: drawing of how the support post fits between the EL meshes. Bottom right: the deflection in $x$ and $y$ at 14 kV and at the best-fit tension and R$_0$ values. The extracted tension amounts between each measurement set are consistent within 30 N. The solid lines represent fits to the experimental data using Eqn.~\ref{eq:deflection} and \ref{eq:def_post}. Measurement of the deflection at the center of the mesh with the post is shifted by 1.5 cm to account for the post.  }
\end{figure}

Figure~\ref{fig:def_magic} shows the measurements of the deflection with and without the support post placed at the center of the mesh. Measurements were taken with a partially tensioned mesh with a frame gap of $\sim$1 mm to study the effect of the post on larger deflections. The post is made of HDPE with a diameter of 3 mm and consists of a hemispherical cap with a thin rod extruding from the base that has a diameter smaller than the mesh hexagons as shown in Fig.~\ref{fig:def_magic}. The rod of the cap fits into a cylindrical tube located between the meshes. The cap and tube tight fitting creates a vacuum seal that holds the post in place between the meshes. The choice of material for the support post was informed by previous studies (see Ref.~\cite{NEXT:2018wtg}) of the breakdown potential of various plastics. These data are fit to determine the mesh tensions that are shown in the legend of Fig.~\ref{fig:def_magic}. The tension values yielded in each fit with and without the post are consistent within 30 N. In addition, the behavior of the support post agrees well with the model where the max deflection is now shifted towards a radius between the center of the mesh and the edge at around 20 cm. The effective post diameter was also left as a free parameter with the support post data and yields a value of 4.6$\pm$0.1~mm, which is 1.4~mm larger than the physical extent of the post. Overall, we find the maximum deflection is reduced by 0.2 mm with the post in place at 14 kV, therefore, the post was not used in NEXT-100. 

\section{EL electric field}\label{appdix:electric_field}
\begin{figure}[t]
\centering
\includegraphics[width=\textwidth]{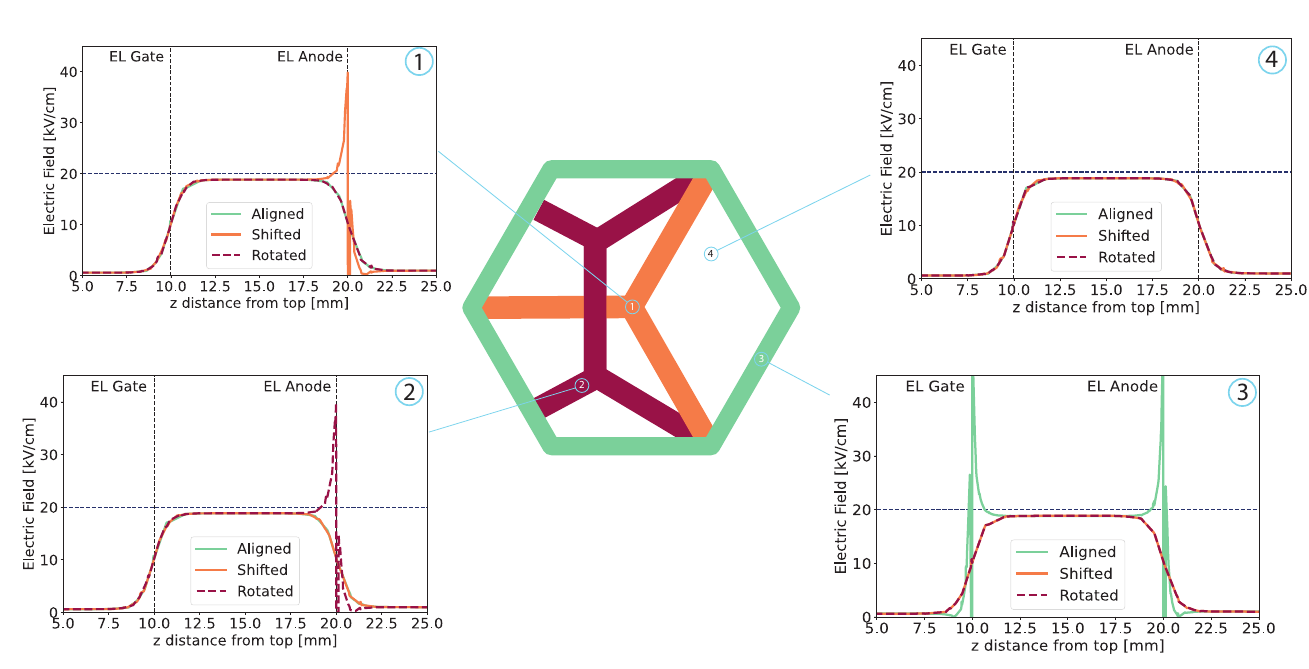}
\caption{\label{fig:z_field} The electric field lines in $z$ at various positions in each configuration. The horizontal blue line shows the electric field for an ideal parallel plate of 20 kV/cm, and each black vertical line shows the position of the gate and anode mesh. One example of a rotated mesh is given in red and the shifted geometry is given in orange.}
\end{figure}

The electric field in the $z$-direction at various starting positions (labeled 1 -- 4) starting from above the gate mesh is shown in Fig.~\ref{fig:z_field} for each mesh configuration. As described in Sec.~\ref{sec:comsol}, the maximum electric field in the region between the gate and anode does not reach 20 kV/cm but a value $\sim$6\% below this. In case 1 at the center of the hexagon cell, the electric field gradually ramps up and down as it crosses the gate and anode region with the field extending beyond the bounds of the mesh. In the case of the shifted geometry, there is a wire land precisely at the center. Towards the surface of the wire, the electric field rapidly rises and then reaches 0 V/cm inside the wire. On the opposite side of the anode wire, we also get a sharp rise in the field, although its magnitude is smaller compared to the side in the high-field region. In case 2, the shifted and rotated fields swap from case 1. In case 3, the field lines penetrate the aligned mesh at the gate and anode so there is a sharp rise and fall after crossing both mesh surfaces. Finally, in case 4 the field shape is symmetrical in all configurations as no wire surface is crossed. 

\FloatBarrier

\bibliographystyle{JHEP}
\bibliography{Bibliography}

\providecommand{\href}[2]{#2}\begingroup\raggedright\begin{thebibliography}{10}

\bibitem{snowmasswhitepaper}
{Adams, C. and others}, {\it Neutrinoless double beta decay},
  \href{http://arxiv.org/abs/2212.11099}{{\tt arXiv:2212.11099}}.

\bibitem{NEXT:2018rgj}
{\bf NEXT} Collaboration, F.~Monrabal et~al., {\it {The Next White (NEW)
  Detector}},  {\em JINST} {\bf 13} (2018), no.~12 P12010,
  [\href{http://arxiv.org/abs/1804.02409}{{\tt arXiv:1804.02409}}].

\bibitem{NEXT:2019qbo}
{\bf NEXT} Collaboration, J.~Renner et~al., {\it {Energy calibration of the
  {NEXT}-White detector with 1\% resolution near Q$_{\beta\beta}$ of
  $^{136}$Xe}},  {\em J. High Energ. Phys.} {\bf 10} (2019) 230,
  [\href{http://arxiv.org/abs/1905.13110}{{\tt arXiv:1905.13110}}].

\bibitem{NEXT:2021dqj}
{\bf NEXT} Collaboration, P.~Novella et~al., {\it {Measurement of the
  $^{136}$Xe two-neutrino double beta decay half-life via direct background
  subtraction in {NEXT}}},  {\em Phys. Rev. C} {\bf 105} (May, 2022)
  [\href{http://arxiv.org/abs/2111.11091}{{\tt arXiv:2111.11091}}].

\bibitem{NEXT:2023jsn}
{\bf NEXT} Collaboration, {C. Adams and others}, {\it Demonstration of
  neutrinoless double beta decay searches in gaseous xenon with next},
  \href{http://arxiv.org/abs/2305.09435}{{\tt arXiv:2305.09435}}.

\bibitem{next100CDR}
{\bf NEXT} Collaboration, {Álvarez, V. and others}, {\it The {NEXT}-100
  experiment for neutrinoless double beta decay searches ({Conceptual Design
  Report})},  \href{http://arxiv.org/abs/1106.3630}{{\tt arXiv:1106.3630}}.

\bibitem{Martin-Albo:2015rhw}
{\bf NEXT} Collaboration, J.~Martin-Albo et~al., {\it {Sensitivity of
  {NEXT}-100 to Neutrinoless Double Beta Decay}},  {\em J. High Energ. Phys.}
  {\bf 05} (2016) 159, [\href{http://arxiv.org/abs/1511.09246}{{\tt
  arXiv:1511.09246}}].

\bibitem{dickinson2014comsol}
E.~J. Dickinson, H.~Ekstr{\"o}m, and E.~Fontes, {\it Comsol
  multiphysics{\textregistered}: Finite element software for electrochemical
  analysis. a mini-review},  {\em Electrochem. Commun.} {\bf 40} (2014) 71--74.

\bibitem{garfieldplusplus}
``{Garfield++}: A toolkit for the simulation of particle detectors.''
  \url{https://garfieldpp.web.cern.ch/}.

\bibitem{Byrnes_2023}
{\bf {NEXT}} Collaboration, N.~Byrnes, I.~Parmaksiz, et~al., {\it {NEXT-CRAB-0:
  a high pressure gaseous xenon time projection chamber with a direct VUV
  camera based readout}},  {\em JINST} {\bf 18} (Aug, 2023) P08006,
  [\href{http://arxiv.org/abs/2304.06091}{{\tt arXiv:2304.06091}}].

\bibitem{baudis2020first}
L.~Baudis, Y.~Biondi, M.~Galloway, and et~al., {\it {The First Dual-Phase Xenon
  TPC Equipped with Silicon Photomultipliers and Characterisation with
  $^{37}\text{Ar}$}},  {\em European Physical Journal C} {\bf 80} (2020) 477,
  [\href{http://arxiv.org/abs/2003.01731}{{\tt arXiv:2003.01731}}].

\bibitem{Stephenson_2015}
S.~Stephenson, J.~Haefner, Q.~Lin, et~al., {\it {MiX: a position sensitive
  dual-phase liquid xenon detector}},  {\em JINST} {\bf 10} (Oct, 2015) P10040,
  [\href{http://arxiv.org/abs/1507.01310}{{\tt arXiv:1507.01310}}].

\bibitem{Baur_2023}
D.~Baur et~al., {\it {The XeBRA platform for liquid xenon time projection
  chamber development}},  {\em JINST} {\bf 18} (Feb, 2023) T02004,
  [\href{http://arxiv.org/abs/2208.14815}{{\tt arXiv:2208.14815}}].

\bibitem{PhysRevD.99.103024}
J.~Xu, S.~Pereverzev, B.~Lenardo, J.~Kingston, D.~Naim, A.~Bernstein,
  K.~Kazkaz, and M.~Tripathi, {\it Electron extraction efficiency study for
  dual-phase xenon dark matter experiments},  {\em Phys. Rev. D} {\bf 99} (May,
  2019) 103024, [\href{http://arxiv.org/abs/1904.02885}{{\tt
  arXiv:1904.02885}}].

\bibitem{Lin_2014}
Q.~Lin, Y.~Wei, J.~Bao, J.~Hu, X.~Li, W.~Lorenzon, K.~Ni, M.~Schubnell,
  M.~Xiao, and X.~Xiao, {\it High resolution gamma ray detection in a dual
  phase xenon time projection chamber},  {\em JINST} {\bf 9} (Apr, 2014)
  P04014, [\href{http://arxiv.org/abs/1309.5561}{{\tt arXiv:1309.5561}}].

\bibitem{APRILE2012573}
{\bf XENON100} Collaboration, E.~Aprile et~al., {\it {The XENON100 dark matter
  experiment}},  {\em Astroparticle Physics} {\bf 35} (2012), no.~9 573--590,
  [\href{http://arxiv.org/abs/1107.2155}{{\tt arXiv:1107.2155}}].

\bibitem{jorg2022characterization}
F.~Jörg, D.~Cichon, G.~Eurin, and et~al., {\it {Characterization of Alpha and
  Beta Interactions in Liquid Xenon}},  {\em European Physical Journal C} {\bf
  82} (2022) 361, [\href{http://arxiv.org/abs/2109.13735}{{\tt
  arXiv:2109.13735}}].

\bibitem{Jewell_2018}
{\bf nEXO} Collaboration, M.~Jewell et~al., {\it {Characterization of an
  Ionization Readout Tile for nEXO}},  {\em JINST} {\bf 13} (Jan, 2018) P01006,
  [\href{http://arxiv.org/abs/1710.05109}{{\tt arXiv:1710.05109}}].

\bibitem{haselschwardt2023measurement}
S.~J. Haselschwardt, R.~Gibbons, H.~Chen, S.~Kravitz, A.~Manalaysay, Q.~Xia,
  P.~Sorensen, and W.~H. Lippincott, {\it {First measurement of discrimination
  between helium and electron recoils in liquid xenon for low-mass dark matter
  searches}},  \href{http://arxiv.org/abs/2308.02430}{{\tt arXiv:2308.02430}}.

\bibitem{Baudis_2021}
L.~Baudis et~al., {\it {Design and construction of Xenoscope — a full-scale
  vertical demonstrator for the DARWIN observatory}},  {\em JINST} {\bf 16}
  (Aug, 2021) P08052, [\href{http://arxiv.org/abs/2105.13829}{{\tt
  arXiv:2105.13829}}].

\bibitem{pcmproducts}
{PCM Products Inc.}, ``Photo chemical etching \& machining.''
  \url{https://www.pcmproducts.com}, 2023.

\bibitem{MakeItFrom.com_2020}
{MakeItFrom.com}, ``{AISI 316Ti (S31635) Stainless Steel}.''
  \url{https://www.makeitfrom.com/material-properties/AISI-316Ti-S31635-Stainless-Steel},
  May, 2020.

\bibitem{SiliconBz}
{Christian, C.}, ``{All About Silicon Bronze - Strength, Properties, and
  Uses}.''
  \url{https://www.thomasnet.com/articles/metals-metal-products/all-about-silicon-bronze/},
  2023.

\bibitem{DiazLopez:2023xxq}
G.~D\'\i{}az~L\'opez, {\em {Sensitivity of NEXT-100 detector to neutrinoless
  double beta decay}}.
\newblock PhD thesis, U. Santiago de Compostela (main), 2023.

\bibitem{LAFERRIERE201593}
B.~LaFerriere, T.~Maiti, I.~Arnquist, and E.~Hoppe, {\it {A novel assay method
  for the trace determination of Th and U in copper and lead using inductively
  coupled plasma mass spectrometry}},  {\em Nucl. Instr. and Meth. in Phys.
  Res. Sect. A} {\bf 775} (2015) 93--98.

\bibitem{VAlvarez_2013}
{\bf NEXT} Collaboration, V.~Álvarez et~al., {\it {Radiopurity control in the
  NEXT-100 double beta decay experiment: procedures and initial measurements}},
   {\em JINST} {\bf 8} (Jan, 2013) T01002,
  [\href{http://arxiv.org/abs/1211.3961}{{\tt arXiv:1211.3961}}].

\bibitem{LEONARD2008490}
D.~Leonard et~al., {\it {Systematic study of trace radioactive impurities in
  candidate construction materials for EXO-200}},  {\em Nucl. Inst. and Meth.
  in Phys. Res. Sect. A} {\bf 591} (2008), no.~3 490--509,
  [\href{http://arxiv.org/abs/0709.4524}{{\tt arXiv:0709.4524}}].

\bibitem{Radiopurity}
PNNL and SNOLAB, ``https://www.radiopurity.org.''
\newblock Aug 2023.

\bibitem{eurofins}
{Eurofins}, ``Eag laboratories.'' \url{https://www.eag.com}, 2023.

\bibitem{akerib2020luxzeplin}
{\bf LUX-ZEPLIN} Collaboration, D.~S. Akerib, C.~W. Akerlof, D.~Y. Akimov,
  et~al., {\it {The {LUX-ZEPLIN} ({LZ}) Radioactivity and Cleanliness Control
  Programs}},  {\em European Physical Journal C} {\bf 80} (2020) 1044,
  [\href{http://arxiv.org/abs/2006.02506}{{\tt arXiv:2006.02506}}].

\bibitem{ALATOUM2020107357}
B.~{Al Atoum}, S.~Biagi, D.~González-Díaz, B.~Jones, and A.~McDonald, {\it
  Electron transport in gaseous detectors with a python-based monte carlo
  simulation code},  {\em Computer Physics Communications} {\bf 254} (2020)
  107357, [\href{http://arxiv.org/abs/1910.06983}{{\tt arXiv:1910.06983}}].

\bibitem{NEXT:2018wtg}
{\bf NEXT} Collaboration, L.~Rogers et~al., {\it High voltage insulation and
  gas absorption of polymers in high pressure argon and xenon gases},  {\em
  JINST} {\bf 13} (2018), no.~10 P10002,
  [\href{http://arxiv.org/abs/1804.04116}{{\tt arXiv:1804.04116}}].

\bibitem{LINEHAN2022165955}
{\bf LUX-ZEPLIN} Collaboration, R.~Linehan et~al., {\it {Design and production
  of the high voltage electrode grids and electron extraction region for the LZ
  dual-phase xenon time projection chamber}},  {\em Nucl. Instr. and Meth. in
  Phys. Res. Sect. A} {\bf 1031} (2022) 165955,
  [\href{http://arxiv.org/abs/2106.06622}{{\tt arXiv:2106.06622}}].

\bibitem{martinez2018calibration}
{\bf NEXT} Collaboration, G.~Mart{\'\i}nez-Lema, J.~H. Morata, B.~Palmeiro,
  A.~Botas, P.~Ferrario, F.~Monrabal, A.~Laing, J.~Renner, A.~Sim{\'o}n,
  A.~Para, et~al., {\it {Calibration of the NEXT-White detector using
  $^{83m}$Kr decays}},  {\em JINST} {\bf 13} (2018), no.~10 P10014,
  [\href{http://arxiv.org/abs/1804.01780}{{\tt arXiv:1804.01780}}].

\bibitem{NYGREN2009337}
D.~Nygren, {\it High-pressure xenon gas electroluminescent tpc for 0-$\nu$
  $\beta\beta$-decay search},  {\em Nucl. Instr. and Meth.} {\bf 603} (2009),
  no.~3 337--348.

\bibitem{CMB_Monteiro_2007}
C.~M.~B. Monteiro, L.~M.~P. Fernandes, J.~A.~M. Lopes, L.~C.~C. Coelho, J.~F.
  C.~A. Veloso, J.~M.~F. dos Santos, K.~Giboni, and E.~Aprile, {\it Secondary
  scintillation yield in pure xenon},  {\em JINST} {\bf 2} (May, 2007) P05001.

\bibitem{solidworks}
{Dassault Systèmes}, {\it Solidworks},  2023.

\bibitem{magboltz}
S.~F. Biagi, {\it Monte carlo simulation of electron drift and diffusion in
  counting gases under the influence of electric and magnetic fields},  {\em
  Nucl. Inst. and Meth. in Phys. Res. Sect. A} {\bf 421} (1999), no.~1-2
  234--240.

\bibitem{SUZUKI1979197}
M.~Suzuki and K.~S., {\it {Mechanism of proportional scintillation in argon,
  krypton and xenon}},  {\em Nucl. Inst. and Meth.} {\bf 164} (1979), no.~1
  197--199.

\bibitem{ALLISON2016186}
J.~Allison et~al., {\it Recent developments in geant4},  {\em Nucl. Inst. and
  Meth. in Phys. Res. Sect. A} {\bf 835} (2016) 186--225.

\bibitem{1610988}
J.~Allison et~al., {\it Geant4 developments and applications},  {\em IEEE
  Trans. Nucl. Sci.} {\bf 53} (2006), no.~1 270--278.

\bibitem{AGOSTINELLI2003250}
S.~Agostinelli et~al., {\it Geant4—a simulation toolkit},  {\em Nucl. Inst.
  and Meth. in Phys. Res. Sect. A} {\bf 506} (2003), no.~3 250--303.

\bibitem{Norman_2022}
L.~Norman, K.~Silva, B.~Jones, et~al., {\it Dielectric strength of noble and
  quenched gases for high pressure time projection chambers.},  {\em Eur. Phys.
  J. C} {\bf 82} (Jan, 2022) [\href{http://arxiv.org/abs/2107.07521}{{\tt
  arXiv:2107.07521}}].

\end{thebibliography}\endgroup

\end{document}